\documentclass[a4paper,fleqn]{cas-dc}

\usepackage[numbers, sort&compress]{natbib}
\usepackage{cleveref}
\usepackage{siunitx}
\usepackage[printonlyused]{acronym}
\usepackage{subcaption}
\usepackage{algorithm}
\usepackage{algorithmic}
\usepackage{nomencl}
\makenomenclature

\usepackage{ifthen}
\renewcommand{\nomgroup}[1]{%
  \ifthenelse{\equal{#1}{P}}{\item[\textbf{Parameters}]}{%
  \ifthenelse{\equal{#1}{V}}{\item[\textbf{Variables / Expressions}]}{%
  \ifthenelse{\equal{#1}{S}}{\item[\textbf{Subscripts}]}{}}}%
}

\hypersetup{
  linkcolor=black,
  urlcolor=black,
  citecolor=black,
  filecolor=black,
  menucolor=black
}
\def\tsc#1{\csdef{#1}{\textsc{\lowercase{#1}}\xspace}}
\tsc{WGM}
\tsc{QE}

\begin{document}
\let\WriteBookmarks\relax
\def\floatpagepagefraction{1}
\def\textpagefraction{.001}

\shorttitle{}    

\shortauthors{W. Liu et~al.}  

\title [mode = title]{Real-Time Control of Sustainable Data Centers: A Two-Layer Model Predictive Control Framework with Workload Flexibility and Heat Recovery} 



%

\author{Wenyu Liu}[
    orcid=0009-0008-8368-4685
]

\cormark[1]

\ead{wenyu.liu@epfl.ch}


\credit{Conceptualization, Data curation, Methodology, Software, Validation, Visualization, Writing - Original Draft}


\author{Enea Figini}[
    orcid=0000-0001-6270-3367
]

\ead{enea.figini@epfl.ch}


\credit{Conceptualization, Methodology, Software, Supervision, Writing - Review \& Editing}

\author{Mario Paolone}[
    orcid = 0000-0001-7073-9036
]


\ead{mario.paolone@epfl.ch}

\credit{Funding acquisition, Project administration, Resources, Supervision, Writing - Review \& Editing}

\affiliation{organization={Distributed Electrical Systems Laboratory, \'Ecole polytechnique f\'ed\'erale de Lausanne}, city={Lausanne}, country={Switzerland}}

\cortext[1]{Corresponding author}



\begin{abstract}
This paper proposes a two-layer model predictive control (MPC) framework for the real-time operation of data centers integrated with on-site photovoltaic generation, battery energy storage, waste heat recovery, and district heating. The upper layer employs scenario-based stochastic optimization to jointly optimize intraday market participation, workload scheduling, and energy management under uncertainty. The lower layer adopts an adaptive tube-based MPC strategy that compensates short-term disturbances while tracking the dispatch references given by the upper layer. The framework further integrates multi-horizon forecasting to support real-time decision making. Microservice-based simulation studies under representative clear-sky, overcast, and oversupply operating conditions demonstrate that the proposed framework accurately tracks dispatch plans despite fast photovoltaic and workload fluctuations. Compared with single-layer control strategies, the adaptive lower-layer controller substantially reduces real-time dispatch deviations and the associated imbalance costs. In addition, the proposed framework naturally adapts to seasonal operating conditions and responds to carbon-aware operating signals, offering a practical approach for economically efficient, sustainable, and grid-supportive operation of future data centers.
\end{abstract}

\begin{graphicalabstract}
\includegraphics[width=\linewidth]{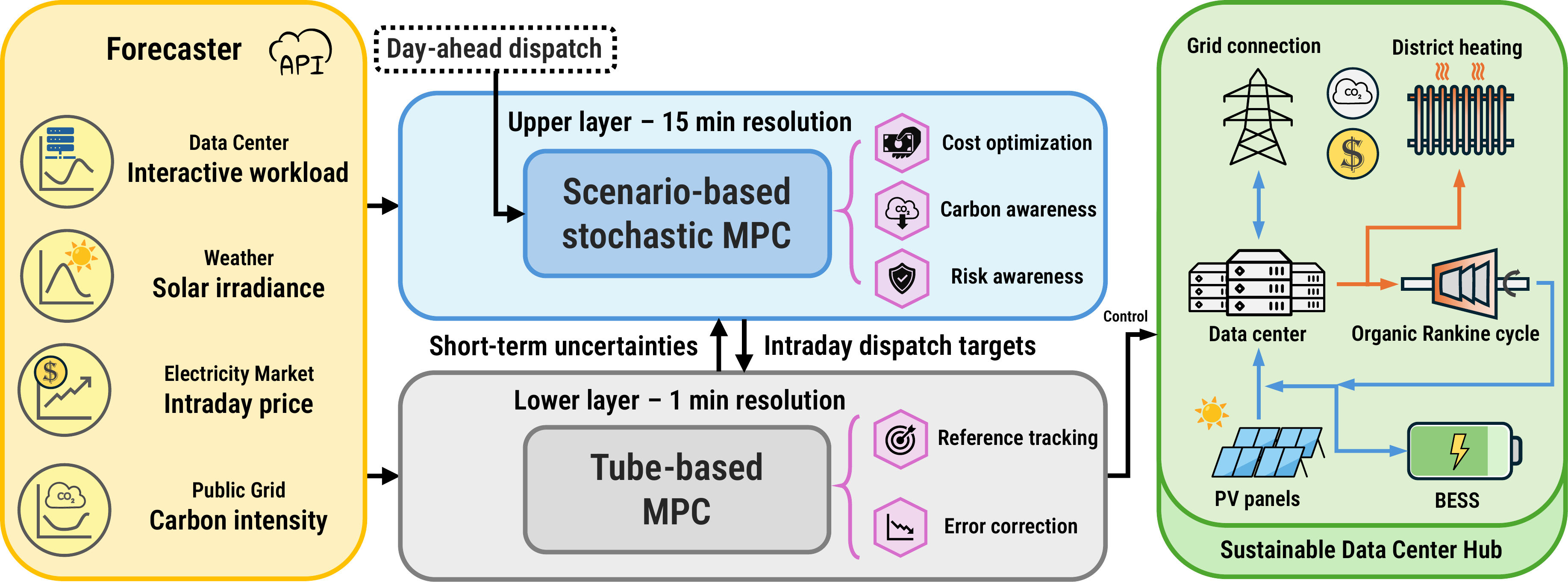}
\end{graphicalabstract}

\begin{highlights}
\item A two-layer MPC framework coordinates data center workloads, DERs, and waste heat recovery.
\item Upper-layer stochastic optimization enables cost- and carbon-aware intraday market dispatch.
\item Lower-layer adaptive tube MPC provides minute-level robust tracking under uncertainty.
\item High-fidelity, microservice-based simulation validates real-time performance.
\end{highlights}



\maketitle


\section*{List of Abbreviations}
\begin{acronym}[DVFS]
    \acro{AI}{Artificial Intelligence}
    \acro{API}{Application Programming Interface}
    \acro{BESS}{Battery Energy Storage System}
    \acro{CI}{Confidence Interval of Uncertainties}
    \acro{CRAC}{Computer Room Air Conditioning}
    \acro{CVaR}{Conditional Value-at-Risk}
    \acro{DA}{Day-Ahead}
    \acro{DER}{Distributed Energy Resource}
    \acro{DH}{District Heating}
    \acro{DtC}{Data Center}
    \acro{DVFS}{Dynamic Voltage and Frequency Scaling}
    \acro{ECDF}{Empirical Cumulative Distribution Function}
    \acro{ECM}{Equivalent Circuit Model}
    \acro{GHI}{Global Horizontal Irradiance}
    \acro{HEX}{Heat Exchanger}
    \acro{ID}{Intraday}
    \acro{IoT}{Internet of Things}
    \acro{MAE}{Mean Absolute Error}
    \acro{MINLP}{Mixed-Integer Nonlinear Programming}
    \acro{MLD}{Mixed Logical Dynamical}
    \acro{MPC}{Model Predictive Control}
    \acro{MTU}{Market Time Unit}
    \acro{ORC}{Organic Rankine Cycle}
    \acro{PCC}{Point of Common Coupling}
    \acro{PI}{Prediction Interval}
    \acro{PUE}{Power Usage Effectiveness}
    \acro{PV}{Photovoltaic}
    \acro{QoS}{Quality of Service}
    \acro{RT}{Real-Time}
    \acro{RES}{Renewable Energy Sources}
    \acro{SLA}{Service-Level Agreement}
    \acro{VCC}{Virtual Capacity Curve}
\end{acronym}

\printnomenclature

\section{Introduction}\label{sec-introduction}

\subsection{Background and motivation} \label{sec-background-and-motivation}

The exponential growth of digital services driven by cloud computing, the \acs{IoT}, big data analytics, and the surge in generative \acs{AI}, has fundamentally transformed how society functions. This digital ecosystem is supported by a global network of \acs{DtC}s. As of the mid-2020s, these facilities have evolved into massive hyperscale complexes operating continuously, processing zettabytes of information on a daily basis \cite{shehabi20242024}. Despite the societal benefits enabled by the digital revolution, its environmental impact is significant. \acs{DtC}s are among the most energy-intensive building types worldwide. It is estimated that they used approximately $\qty{205}{TWh}$ and $\qty{415}{TWh}$ of electricity in 2018 and 2024, respectively, accounting for around $\qty{1}{\percent}$ and $\qty{1.5}{\percent}$ of global electricity consumption in the corresponding years \cite{masanetRecalibratingGlobalData2020, IEA2025EnergyAI}. This demand is projected to more than double by 2030 as \acs{AI}-driven workloads require higher computational intensity and power density \cite{IEA2025EnergyAI}. { In Europe, the recent study \cite{entso-eENTSOEDataCentres2026} estimates an increase of over 50~\% the total electricity demand from \acs{DtC}s between 2025 and 2030.} Beyond electricity consumption alone, the carbon footprint of \acs{DtC}s is also becoming a major concern: despite widespread corporate net-zero commitments, the \acs{DtC} industry is projected to contribute up to $\qty{8}{\percent}$ of global carbon emissions by 2030 \cite{caoSystematicSurveyCarbon2022}. In response, major \acs{DtC} operators have announced ambitious decarbonization targets, such as Google's commitment to operate on 24/7 carbon-free energy by 2030 \cite{google2026environmental} and Amazon's pledge to achieve net-zero carbon emissions across its operations by 2040 \cite{amazon2025sustainability}. These commitments further motivate the development of strategies that explicitly incorporate carbon-aware operation.

At the same time, the hosting capacity of \acs{DtC}s in the electrical grid is approaching a critical point. The rapid and often uncoordinated integration of large-scale \acs{DtC}s poses increasing risks to grid operation, particularly in already congested power systems \cite{kwonOperationalRisksGrid2025}. To address these emerging risks, some countries are discussing to update their national grid connection codes, which means the traditional inelastic power demand behavior of \acs{DtC}s may no longer be allowed in the future networks \cite{entso-eENTSOEDataCentres2026}. Moreover, as electricity grids transition to higher penetrations of stochastic \acs{RES}, there is a growing need for flexible demand that can actively contribute to the grid balancing. \acs{DtC}s may provide such demand-side flexibility if they are forced by grid codes and/or incentivized by market opportunities. Their computational workloads can, to some extent, be shifted temporally or geographically without violating \acs{SLA}s \cite{zhangRemuneratingSpaceTime2022, chenOperationalFlexibilityActive2021}. {Furthermore, to facilitate sustainable operation, \acs{DtC}s are increasingly integrated into multi-energy systems, which incorporate a diverse portfolio of \acs{DER}s, including \acs{PV} arrays, wind turbines, gas turbines, \acs{BESS}s, and \acs{DH} \cite{lyuOptimalSizingEnergy2021, maMultiobjectiveMultistageDistributionally2025, figiniShouldSmallScaleData2026, panUncertaintyawareReinforcementLearning2026}.} Nevertheless, exploiting the flexibility of the system in practice is far from trivial. \acs{RT} coordination must simultaneously account for \acs{QoS} constraints, volatile energy costs, uncertain \acs{RES}, and carbon-aware operational objectives. Addressing these challenges requires advanced control and optimization frameworks that tightly integrate system modeling, power market integration, forecasting, and \acs{RT} decision-making.

\subsection{Problem statement}

As mentioned above, the increasing integration of flexible workloads, \acs{DER}s, and electricity market participation, transforms the operation of modern \acs{DtC}s into a complex multi-timescale decision-making problem. Operational decisions must simultaneously satisfy computational service requirements, coordinate heterogeneous energy resources, and respond to evolving market and environmental signals. A key challenge arises from the different temporal characteristics of these decisions. While energy trading and resource planning require anticipation of future uncertainties over several hours to a whole day, \acs{RT} operation must continuously compensate for forecast errors in \acs{DER}s and workloads. Consequently, achieving economic efficiency, carbon-aware operation, and operational robustness within a unified framework remains difficult. This work therefore addresses the following research question:
\begin{quote}
Given a \acs{DA} operation plan, how can a \acs{DtC}-centric multi-energy system be operated in \acs{RT} such that workload flexibility and energy resources are coordinated to achieve economical, carbon-aware, and robust operation under uncertainty?
\end{quote}
To answer this question, a hierarchical control framework integrating forecasting, market participation, and \acs{RT} optimization is proposed.

\subsection{Literature review} \label{sec-literature-review}

\begin{table*}[ht]
    \centering
    \scriptsize
    \caption{Comparison of this work and related works. $\text{WL}_\text{flex}$, $\pm$, and \acs{RT} Cmp. refer to the \acs{DtC} workload flexibility, uncertainty and real-time disturbance compensation, respectively.}\label{tbl-literature-review}
    \begin{tabular}{ccccccccccccc}
    \toprule
        Ref. & Job model & $\text{WL}_\text{flex}$ & DtC power model & RES & BESS & ORC & CO\textsubscript{2} & Market & Field & $\Delta T$ & $\pm$ & \acs{RT} Cmp.\\ 
    \midrule
        \cite{cupelliDataCenterControl2018} & M/M/1 & $\checkmark$ & High fidelity, air &   & $\checkmark$ &   &   &   & RT scheduling & 1min &  & \\ 
        \cite{fangQoSDrivenPowerManagement2016} & M/M/c &  & Heat dynamics, air &   &   &   &   &   & \acs{DtC} management & 15min/1min &  & \\ 
        \cite{jawadRobustOptimizationTechnique2021} &   &   & Fixed PUE & $\checkmark$ & $\checkmark$ &   &   & DA & DA/ID scheduling & 1h & $\checkmark$ & \\ 
        \cite{chenOperationalFlexibilityActive2021} & M/M/1 & $\checkmark$ & Fixed PUE & $\checkmark$ &   &   &   &   & Flexibility analysis & 15min &  & \\ 
        \cite{lyuOptimalSizingEnergy2021} & M/M/1 &   & Fixed PUE & $\checkmark$ & $\checkmark$ &   &   &   & Optimal sizing & 1h &   \\ 
        \cite{zhuEnergyOptimalDispatch2022} & M/M/1 & $\checkmark$ & Fixed PUE & $\checkmark$ & $\checkmark$ &   &   & DA & DA/ID scheduling & 1h & $\checkmark$ & \\ 
        \cite{radovanovicCarbonAwareComputingDatacenters2023} &   & $\checkmark$ & Fixed PUE &   &   &   & $\checkmark$ & DA & DA scheduling & 1h & $\checkmark$ & \\ 
        \cite{wangHierarchicalDispatchStrategy2023} &   &   &   &   & $\checkmark$ &   &   & DA & UPS dispatch & 1h/15min &  & \\ 
        \cite{guoMultitimescaleOptimizationScheduling2024} & M/M/1 & $\checkmark$ & Linear principle & $\checkmark$ & $\checkmark$ &   &   & DA & DA/ID scheduling & 1h/15min & & \\ 
        \cite{zhouDataCenterLoad2025} & M/M/1 & $\checkmark$ & Heat dynamics, air &   &   &   &   &   & Load modeling &   &  & \\ 
        \cite{xieIntegratingStateDependentQueue2024} & M/M/c &   & Linear principle &   &   &   & $\checkmark$ &   & DA scheduling & 1h &  & \\ 
        \cite{wangCoordinatedOptimizationDistributed2025} &   &   & Explicit & $\checkmark$ & $\checkmark$ &   &   & DA & UPS dispatch & 1h/15min & & \\ 
        \cite{hanDatadrivenDistributionallyRobust2025} & $\rm T_{exec}$ & $\checkmark$ & Linear principle & $\checkmark$ & $\checkmark$ &  & $\checkmark$ & DA & DA/ID scheduling & 1h & $\checkmark$ & \\ 
        \cite{maMultiobjectiveMultistageDistributionally2025} &   & $\checkmark$ & Linear principle & $\checkmark$ & $\checkmark$ & $\checkmark$ & $\checkmark$ &   & Planning & 1h & $\checkmark$ & \\
        \cite{figiniShouldSmallScaleData2026} & & $\checkmark$ & Linear principle & $\checkmark$ & $\checkmark$ & $\checkmark$ & $\checkmark$ & DA & DA scheduling & 1h & $\checkmark$ & \\
        \cite{hallCarbonAwareComputingData2025} & FIFO & $\checkmark$ & Linear principle & & & & $\checkmark$ & DA & DA/RT scheduling  & 1h & $\checkmark$ &  \\
        This & M/M/1 & $\checkmark$ & High fidelity, hybrid & $\checkmark$ & $\checkmark$ & $\checkmark$ & $\checkmark$ & ID & ID/RT control & 15min/1min & $\checkmark$ & $\checkmark$ \\
        \bottomrule
    \end{tabular}
\end{table*}

Research on \acs{DtC} operational control strategies spans multiple disciplines. This section reviews the strands of literature most relevant to the scope of this work.

\subsubsection{Data center modeling}

Significant research effort has been devoted to modeling \acs{DtC}s from two perspectives:
\begin{itemize}
    \item workload flexibility and execution dynamics, and
    \item the relationship between workload execution and power consumption.
\end{itemize}

\paragraph{Workload modeling}
Most studies classify \acs{DtC} workloads into two categories: \emph{interactive} and \emph{batch} workloads. Interactive workloads (e.g., \acs{AI} interfaces, web services, SaaS, transactions \cite{entso-eENTSOEDataCentres2026}) must be served immediately upon arrival to satisfy strict latency requirements, whereas batch workloads (e.g., backups, \acs{AI} training \cite{entso-eENTSOEDataCentres2026}) can be deferred and processed flexibly as long as predefined deadlines are met \cite{chenOperationalFlexibilityActive2021, radovanovicCarbonAwareComputingDatacenters2023, zhuEnergyOptimalDispatch2022}. To ensure \acs{DtC} \acs{QoS} constraints, several works adopt queuing-theoretic formulations, most commonly M/M/1 queue models\footnote{{ M/M/1 queue models represent the length of the queue in a single server system where job arrivals follow a Poisson process and execution times follow
an exponential distribution.}}, and typically assume homogeneous workloads and hardware resources \cite{chenMultiobjectiveRobustOptimal2023, chenOperationalFlexibilityActive2021}. While these assumptions enable tractable analytical formulations of the workload, they may limit applicability in heterogeneous and large-scale \acs{DtC} environments. In contrast, other studies focus on more detailed and fine-grained workload execution models, designing online job scheduling algorithms that explicitly track individual job arrivals, deadlines, and execution states \cite{liuOnlineJobScheduling2023}. Although such approaches improve modeling fidelity, they often increase computational complexity, making \acs{RT} integration with \acs{DtC}-level energy management more challenging.

\paragraph{Power modeling} Accurately mapping \acs{DtC} resource (utilization such as CPU, memory, and server configurations) to electrical power consumption is essential for control and optimization frameworks. \cite{radovanovicPowerModelingEffective2022} proposed both a piecewise linear model and a nonlinear unified model based on random forests to relate CPU utilization across Google’s power domains to power consumption. These models have been deployed in production environments to enable cost- and carbon-aware load management and power provisioning. Simpler linear power models are also widely adopted. For instance, \cite{guoMultitimescaleOptimizationScheduling2024} models computing power as a linear function of workload processing capacity, added on top of a constant idle power component. Due to their simplicity, such models offer favorable computational properties and are particularly suitable for \acs{RT} optimization. At the hardware level, power management techniques such as \acs{DVFS} have been extensively studied as a means to improve energy efficiency \cite{lesueurDynamicVoltageFrequency2010a}. Building on these concepts, \cite{wanJointCoolingServer2018} jointly considers processor-level power management and \acs{DtC} cooling systems, proposing a cross-layer optimization framework for holistic \acs{DtC} energy minimization that collaborates different components and dynamically configures parameters according to the workload condition. However, the resulting formulation is a \acs{MINLP} problem, well known to be computationally intensive for \acs{RT} control applications. To address this limitation, recent work \cite{zhouDataCenterLoad2025} extended the framework of \cite{wanJointCoolingServer2018} by proposing a hierarchical load modeling approach. A fine-grained foundational layer captures the electrical–thermal–performance interactions, while the overall \acs{DtC} power consumption is represented as a nonlinear function of workload execution and linearized in a piecewise manner. 

\subsubsection{Interaction with the electricity market}

Recent studies have investigated how \acs{DtC}s can participate in electricity markets through optimal bidding and scheduling strategies. For example, \cite{chenMultiobjectiveRobustOptimal2023} proposed an optimal bidding framework for a \acs{DtC} operator participating in the \acs{DA} electricity market. \cite{figiniShouldSmallScaleData2026} proposed a carbon-aware and cost-effective framework for small scale \acs{DtC}s to integrate into the \acs{DA} electricity market, which shows a cost reduction potential. \cite{zhangRemuneratingSpaceTime2022} analyzed remuneration mechanisms for \acs{DtC} load-shifting flexibility in both temporal and geographical dimensions within market clearing schemes, while \cite{caoManagingDataCenter2024} explored the potential role of geo-distributed \acs{DtC} in the balancing market. In addition, \cite{cupelliDataCenterControl2018} presented a framework for the internal energy management of \acs{DtC}s equipped with \acs{BESS} for participating in demand response programs, while the on-site \acs{RES} is not modeled. However, these studies primarily focus on \acs{DA}, balancing, or demand response markets and largely overlook participation in electricity markets within \acs{RT} operational settings.

\subsubsection{Control strategy}

As summarized in \Cref{tbl-literature-review}, a variety of operational control strategies have been proposed for sustainable \acs{DtC}s equipped with on-site \acs{RES} and \acs{BESS}. One important research direction focuses on exploiting workload flexibility. For example, \cite{radovanovicCarbonAwareComputingDatacenters2023} proposed Google's carbon-intelligent computing system, where the \acs{VCC} computed in the \acs{DA} stage constrains the computing resources available for \acs{RT} job scheduling. Building upon this concept, \cite{hallCarbonAwareComputingData2025} integrated temporal and spatial workload flexibility with \acs{VCC} using a distributionally robust optimization framework consisting of a \acs{DA} planner and a \acs{RT} job placement module. Another line of research investigates the coordination between \acs{DtC}s and energy systems. \cite{michaelEconomicSchedulingVirtual2023} modeled the battery system of a \acs{DtC} as a virtual power plant to optimize participation in both \acs{DA} and \acs{RT} electricity markets, although workload flexibility was not considered. Multi-timescale dispatch strategies have also been developed by \cite{zhuEnergyOptimalDispatch2022} and \cite{guoMultitimescaleOptimizationScheduling2024}, where stochastic \acs{MPC} is employed to coordinate \acs{DA} scheduling with \acs{ID} dispatch corrections. {However, the control time resolution remains at \qty{15}{min} (or above), which is adequate for scheduling but insufficient for compensating fast disturbances such as \acs{RES} fluctuations, and workload variability that occur within a \acs{MTU}.} In parallel, \cite{wangHierarchicalDispatchStrategy2023,wangCoordinatedOptimizationDistributed2025} investigated hierarchical control of \acs{DtC} uninterruptible power supply systems, but without explicitly exploiting the flexibility of computing workloads. Overall, existing studies have demonstrated the value of workload flexibility, hierarchical energy management, and multi-timescale optimization. Nevertheless, few have considered a unified framework that combines \acs{DA} scheduling, \acs{ID} market participation, and minute-level robust \acs{RT} control while jointly coordinating flexible workloads and multiple on-site \acs{DER}s.

{
\subsubsection{Research gap and positioning}

The literature reviewed above provides important building blocks for sustainable \acs{DtC} operation, including workload flexibility, energy resource coordination, electricity market participation, and multi-timescale control. However, these elements have rarely been integrated across their distinct operational timescales (from \acs{DA} to minute-level \acs{RT} operation). Existing hierarchical \acs{MPC} studies in power systems \cite{raimondicominesiTwoLayerStochasticModel2018, guptaReliableDispatchActive2022} demonstrate the effectiveness of separating scheduling and \acs{RT} control, but do not account for the distinctive flexibility of \acs{DtC} computing workloads together with \acs{ID} market adjustments. Conversely, \acs{DtC}-oriented studies typically consider workload flexibility and market participation, but often retain a hourly or slower control resolution. Overall, existing studies have demonstrated the value of workload flexibility, hierarchical energy management, and multi-timescale optimization. However, they differ substantially in the temporal resolution, controllable resources, and market mechanisms considered, leaving their joint coordination in \acs{DtC}-centric multi-energy systems insufficiently explored.

}

\subsection{Contribution of this work}

The main contributions of this work are summarized as follows:

{
\begin{itemize}
    \item A multi-timescale energy management framework for sustainable \acs{DtC}s that jointly coordinates workload flexibility, electricity market participation, and on-site energy resources, using stochastic quarter-hourly scheduling and minute-level robust \acs{RT} control.

    \item An integrated optimization formulation that captures the coupled operation of flexible workloads, \acs{BESS}, \acs{PV}, low-entropic waste heat recovery, and \acs{DH}, enabling cost- and carbon-aware operation under uncertainty.

    \item A practical hierarchical control implementation in which the upper layer establishes economically efficient operating targets while the lower layer provides fast disturbance compensation using adaptive uncertainty bounds and available energy flexibility.
    
    \item  A comprehensive validation study under different seasonal and market conditions, demonstrating the ability of the proposed framework to coordinate market decisions and heterogeneous energy resources while maintaining accurate real-time operation under disturbances.
\end{itemize}}

\section{System model}\label{sec-system-model}
\begin{figure}[pos=htbp]
    \centering
    \includegraphics[width=0.95\linewidth]{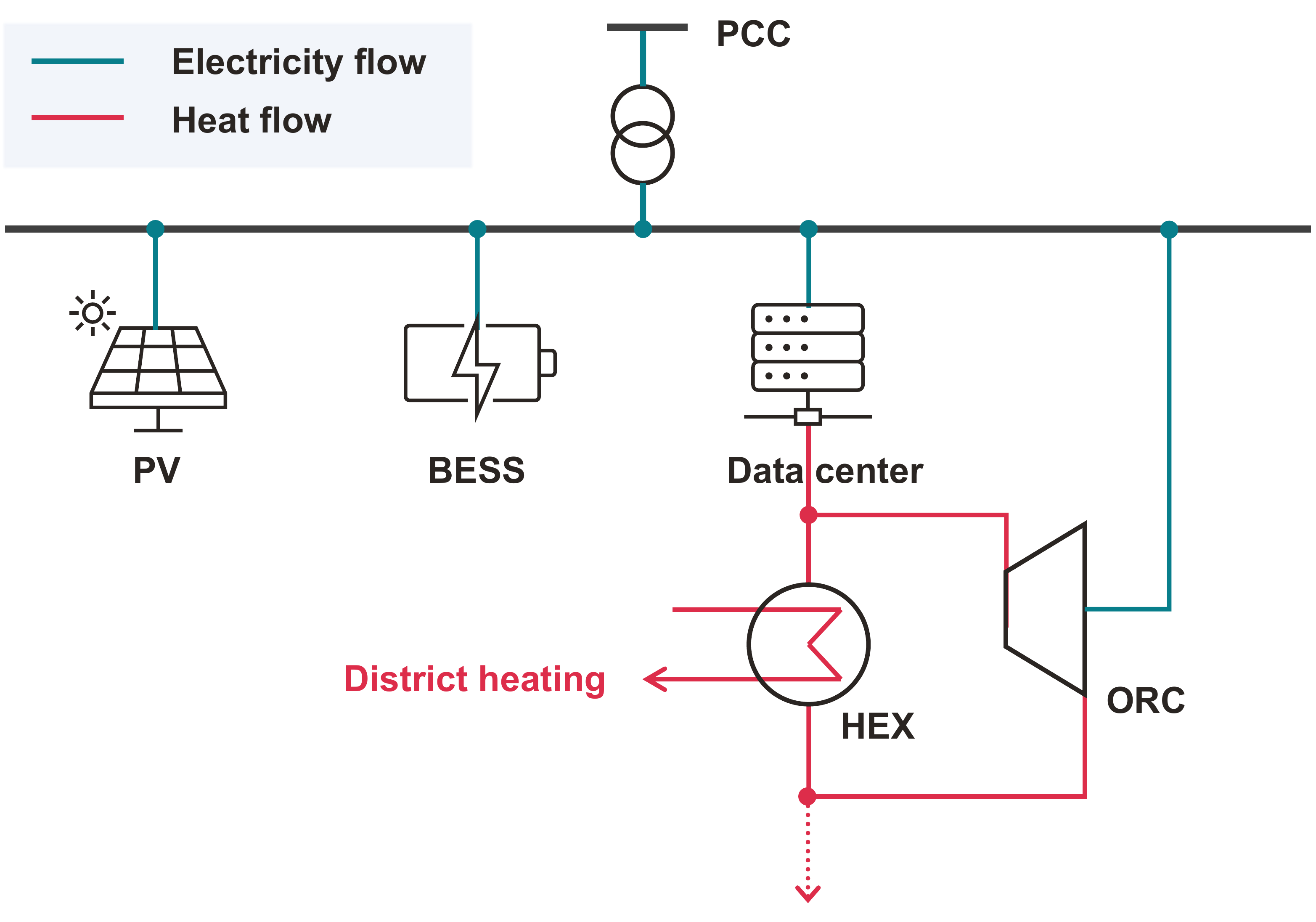}
    \caption{The single line diagram of the system.}
    \label{fig-SLD}
\end{figure}

{The considered system configuration, as illustrated in \Cref{fig-SLD}, is representative of emerging sustainable \acs{DtC}s, where on-site \acs{RES} generation, \acs{BESS}, and waste heat recovery are deployed to improve both economic and environmental performance. This configuration is consistent with the experimental testbed developed in the Heating Bits project \cite{HeatingBits} and serves as a representative platform for evaluating the proposed control framework.} 
The system is connected to the external utility grid at the \acs{PCC}. The electrical infrastructure includes the following key assets:
\begin{itemize}
    \item \acs{DtC}: the core user of the electricity within the system, has a limited flexibility from the shiftable workload.
    \item \acs{PV} plant: provides local renewable generation to reduce dependency on the grid and potentially reduces the carbon footprint.
    \item \acs{BESS}: acts as the core controllable resource, provides flexibility to the system.
    \item \acs{ORC} regenerator: a recovery unit that converts a portion of the medium grade waste heat from the \acs{DtC} back into electricity from \acs{ORC} to save energy.
\end{itemize}
Additionally, the heat loop is processed through two main channels:
\begin{itemize}
    \item \acs{DH}: uses an \acs{HEX} to valorize the waste heat by transferring it in the local district heating network.
    \item \acs{ORC} turbine: use the temperature differential to drive a turbine for regenerating electricity.
\end{itemize}
To design a control framework for the aforementioned system, the behaviors and technical constraints of each component are characterized in the following sections.

\subsection{Data center}
The \acs{DtC} in this system is not an inelastic electrical load but functions as a partly and temporally flexible load. In other words, it has the ability to shift batch computing tasks in time without violating \acs{SLA}s \cite{radovanovicCarbonAwareComputingDatacenters2023}. In this work, the method proposed by \cite{zhouDataCenterLoad2025} is used to model the \acs{DtC} through optimal holistic energy consumption characteristics, which considers both the interactive workload and the batch workload, as well as the characteristics of the equipments. {It is assumed that there is a built-in optimizer for the \acs{DtC} internal energy management, which coordinates the server-level IT resources and cooling infrastructure to minimize energy consumption for a given set of tasks. The built-in optimizer receives workload execution targets together with the current operating conditions (e.g., ambient temperature and active server count), and determines the corresponding server allocation, processor frequencies, cooling operation, and other internal control actions to minimize the total facility power consumption while satisfying \acs{QoS} constraints. These internal decisions are abstracted from the \acs{MPC} framework, which only observes the resulting aggregate power consumption through the \acs{DtC} power model.} From an external system perspective, the total power consumption of the \acs{DtC}, $p_\text{dtc}$, can be expressed as a function of the execution rates of these two workload types, shown in \eqref{eq-p-dtc}:
\begin{equation}
   p_\text{dtc} = \varphi(a,b) \label{eq-p-dtc}
\end{equation}
where $\varphi$ represents the implicit non-linear mapping of workload execution to total facility power given by the \acs{DtC} built-in optimizer, $a$ and $b$, respectively, represent the execution rates in $\unit{unit/s}$ of interactive workloads (must be executed immediately after arrival) and batch workloads (need to be completed within a certain duration, e.g., 24 hours) \cite{zhouDataCenterLoad2025}.

\nomenclature[V]{$a, b$}{The execution rate of interactive and batch workloads}
\nomenclature[V]{$\varphi, \hat{\varphi}$}{The implicit non-linear mapping of workload execution to total \acs{DtC} power and its explicit linear fit}

\subsubsection{Workload model}

\paragraph{Queuing model}\label{sec-dtc-queuing-model}

To characterize the relationship between workload execution and service performance, the \acs{DtC} is modeled as a queuing system. Interactive requests are assumed to arrive according to a Poisson stochastic process, while service times are exponentially distributed, which leads to a classical M/M/c formulation \cite{kendallStochasticProcessesOccurring1953}. Although the M/M/c model provides a realistic representation of a parallel server cluster, its response-time expression relies on Erlang’s C formula and becomes computationally expensive when embedded into \acs{RT} optimization problems \cite{gautam2012analysis}. Therefore, following \cite{chenMultiobjectiveRobustOptimal2023}, an independent M/M/1 approximation is adopted.

For a \acs{DtC} with $N$ active servers, the average arrival rate per server is defined by:
\begin{equation}
\lambda = \frac{a + b}{N} \label{eq-dtc-lambda}
\end{equation}
The expected response time, which serves as the \acs{QoS} metric for interactive workloads, is approximated by:
\begin{equation}
t_D = \frac{1}{f \cdot \mu - a/N}
\end{equation}
where $\mu$ is the nominal service rate of a server and $f \in [f_{\min}, f_{\max}]$ is the relative operating frequency. { The independent M/M/1 approximation approximation is introduced solely as a computationally tractable \acs{QoS} constraint for the subsequent \acs{DtC} power-model construction, rather than as a detailed queuing representation of server-level queuing dynamics. Its conservatism over the operating region considered in this study is assessed against the corresponding M/M/c model in Appendix \ref{sec-app-queuing-model-comparison}.}

\nomenclature[P]{$N$}{The number of active servers}
\nomenclature[P]{$t_D$}{The response time of interactive workloads}
\nomenclature[P]{$\mu$}{The nominal service rate of the server}
\nomenclature[V]{$f$}{The relative operating frequency of the server}

\paragraph{Unit model}\label{sec-dtc-unit-model}

To bridge the gap between abstract queuing units and physical hardware utilization, we define the service rate $\mu$ based on the hardware architecture. The relationship between the processor operating frequency $f$, architectural efficiency $\eta$, and job complexity $C$ is established as:
\begin{equation}
\mu = \frac{f \cdot \eta_{\rm processor}}{C}
\end{equation}
where $C$ represents the total arithmetic operations (FLOP) required per unit of work, and $\eta$ represents the throughput (in FLOP/cycle). This formulation allows to translate standard processor hours into an equivalent arrival rate. If a user requests $\chi$ processor hours based on historical usage at frequency $f_{\text{hist}}$, the workload can be represented as
\begin{itemize}
    \item An arrival rate of $\displaystyle \frac{f_{\text{hist}} \cdot \eta}{C}$ sustained over $\chi$ hours.
    \item Alternatively, an arrival rate of $\displaystyle  \chi \cdot \frac{f_{\text{hist}} \cdot \eta}{C}$ normalized to a 1-hour window.
\end{itemize}

\nomenclature[P]{$C$}{The total arithmetic operations required per unit of work}
\nomenclature[P]{$\eta_{\rm processor}$}{The architectural efficiency of the processor}


\paragraph{Workload Dynamics}

Based on the workload characterization introduced before, the execution dynamics of interactive and batch workloads in the \acs{DtC} are modeled by the set of constraints in \eqref{eq-dtc-wl-dynamics}:
\begin{subequations}
\label{eq-dtc-wl-dynamics}
    \begin{align}
    & a^{t} = \mathrm{a}_\text{arr}^{t} \\
& a^{t} + b^{t} \leq \mathrm{WL}_\text{max} \\
& b_\text{job}^{t} \leq b^{t} \label{eq-con-b-job-dtc} \\
& B_\text{executed}^{t+\Delta T} = B_\text{executed}^{t} + \Delta T \cdot b_\text{job}^{t}  \label{con-dtc-b-execution-dynamic}
\end{align}
\end{subequations}
where $t$ denotes the discrete time index and $\Delta T$ is the length of the time step. The variables $a^{t}$ and $b^{t}$ represent the execution rates of interactive and batch workloads, respectively, while $a_\text{arr}^{t}$ denotes the arrival rate of interactive workloads. The total workload executed at each time step is constrained by the maximum processing capacity of the \acs{DtC}, $\mathrm{WL}_\text{max}$. The variable $b_\text{job}^{t}$ represents the effective batch workload execution rate that contributes to serving batch job requests. Constraint~\eqref{eq-con-b-job-dtc} allows the \acs{DtC} to execute a higher batch workload rate $b^{t}$ than the rate required by the job queue, thereby providing operational flexibility; however, only the effective execution rate $b_\text{job}^{t}$ is accumulated in the executed batch workload state $B_\text{executed}^{t}$ through the dynamic update in \eqref{con-dtc-b-execution-dynamic}.

\nomenclature[P]{$\mathrm{a}_\text{arr}$}{The arrival rate of interactive workloads}
\nomenclature[P]{$t$}{The general discrete time index}
\nomenclature[P]{$\Delta T$}{The length of the time step}
\nomenclature[P]{$\mathrm{WL}_\text{max}$}{The maximum processing capacity of the \acs{DtC}}
\nomenclature[V]{$b_\text{job}$}{The effective batch workload execution rate that contributes to servicing batch job requests}
\nomenclature[V]{$B_\text{executed}$}{The executed state of the booked batch workload}

\subsubsection{Power model}

The total power consumption of the \acs{DtC} is a composite of IT equipment demand and the auxiliary power required for the thermal management and other equipments. To accurately model the heat recovery potential, the power consumption is split into liquid-cooled and air-cooled components.

\paragraph{Computational power} We use the load modeling method proposed by \cite{zhouDataCenterLoad2025} to model the ideal power of a server $p_\text{sid}$ governed under \acs{DVFS}, in which the server parameters come from system tests \cite{specSPECPowerHuawei}. To account for the hybrid cooling strategy, this power is partitioned:
\begin{equation}
p_{\text{sid}} = \underbrace{r \cdot p_{\text{sid}}}_{p_{\text{processor}}} + \underbrace{(1-r) \cdot p_{\text{sid}}}_{p_{\text{non-processor}}}
\end{equation}
where $r$ is the fraction of heat dissipated via liquid-cooled cold plates (primarily computing processors), and $(1-r)$ represents air-cooled components (memory, storage, and voltage regulators). To account for the temperature-dependent leakage power, the computing power consumptions are calculated using a linear thermal coefficient $F_{\text{T}}(T) = c_{\text{lk},1}T + c_{\text{lk},2}$ \cite{arrobaLeakagePowerModelingOptimal2013, zhouDataCenterLoad2025}:
\begin{equation}
\begin{aligned}
    & p_{\text{processor, cp}} = F_{\text{T}}(T_{\text{processor}}) \cdot p_{\text{processor}} \\
    & p_{\text{non-processor, cp}} = F_{\text{T}}(T_{\text{non-processor}}) \cdot p_{\text{non-processor}}
\end{aligned}\label{eq-p-x-cp}
\end{equation}

\nomenclature[V]{$p_\text{sid}$}{The ideal power consumption of a server}
\nomenclature[V]{$p_{\text{processor}}$}{The processors' power contribution to $p_\text{sid}$}
\nomenclature[V]{$p_{\text{non-processor}}$}{The non-processor components' power contribution to $p_\text{sid}$}
\nomenclature[P]{$r$}{The fraction of heat dissipated via liquid-cooling and air-cooling}

\paragraph{Air cooling and fan power}
In the \acs{DtC}, the air cooling is performed by fan arrays and a \acs{CRAC} unit. The fan cooling system presented in \cite{zhouDataCenterLoad2025} is used to model the relationship between the supply air temperature $T_\text{SA}$, the rotational speed of the fan array $r_{\rm fan}$, the fan array power $p_{\text{sf}}$, and the power consumed by \acs{CRAC}, $p_{\text{cl, air}}$.
\nomenclature[V]{$r_{\rm fan}$}{The rotational speed of the fan array}
\nomenclature[V]{$T_\text{SA}$}{The supply air temperature of the air-cooling system}
\nomenclature[V]{$p_{\text{sf}}$}{The fan power of a server}
\nomenclature[V]{$p_{\text{cl, air}}$}{The power consumed by \acs{CRAC} for air-cooling}

\paragraph{Liquid cooling and pump power} \label{sec-liquid-cooling-and-pump-power}

High-density heat from computing processors is dissipated via cold plates and sent to a liquid cooling loop. To act as a stable heat source for the waste heat valorization, the inlet temperature $T_{\text{in}}$ and outlet temperature $T_{\text{out}}$ are designed to be maintained stable by modulating the mass flow rate $\dot{m}$.  The heat balance is defined as:
\begin{equation}
p_\text{processor, cp} = \dot{m} \cdot c_\text{p} \cdot (T_\text{out} - T_\text{in})
\end{equation}
where $c_\text{p}$ is the constant pressure specific heat capacity of the coolant. The power required by the pumps, $p_{\text{pump}}$, is proportional to the flow rate and the pressure drop $\Delta P$ \cite{kochupurackalrajanIntegratedSiliconMicrofluidic2022}:
\begin{equation}
p_\text{pump} = \frac{1}{\eta_{\rm pump}}\cdot\frac{\dot{m}}{\rho} \cdot \Delta P  \qquad \Delta P \propto \dot{m}^\beta \qquad \beta \approx 1
\end{equation}
where $\eta_{\rm pump}$ is the pump efficiency and $\rho$ is the density of the fluid. Thus the power consumption of the liquid-cooling system is given by:
\begin{equation}
    p_{\text{cl, liquid}} = N\cdot p_\text{pump}
\end{equation}
From the heat transfer model:
\begin{equation}
p_\text{processor, cp} = (T_\text{out} - T_\text{in})/R_\text{total}
\end{equation}
The thermal resistance $R_\text{total}$ is sensitive to the flow regime, according to \cite{vanerpCodesigningElectronicsMicrofluidics2020}:
\begin{equation}
R_\text{total} = R_\text{const} + \mathrm{C}\cdot \dot{m}^\gamma\qquad \gamma\approx -1
\end{equation}

\nomenclature[P]{$T_\text{in}, T_\text{out}$}{The inlet and outlet temperature of the liquid cooling system}
\nomenclature[P]{$c_\text{p}$}{The specific heat capacity of the coolant}
\nomenclature[V]{$\dot{m}$}{The mass flow rate of the \acs{DtC} liquid coolant}
\nomenclature[P]{$\eta_{\{ \cdot\}}$}{The efficiency of the component or process $\{ \cdot\}$}
\nomenclature[P]{$\rho$}{The density of the coolant fluid}
\nomenclature[V]{$p_\text{processor, cp}$}{The computing power of processors}
\nomenclature[V]{$R_\text{const}, R_\text{total}$}{The constant and total thermal resistance}

\paragraph{Other equipments}

\acs{DtC}s are usually equipped with units {such as UPS, backup chillers, and redundant nodes,} here we include a simple constant to model the standby consumptions $p_\text{standby}$. 

\paragraph{Total facility power}
The aggregate power consumption of the \acs{DtC}, $p_{\text{dtc, obj}}$, used as the objective function of the built-in optimizer is given by \eqref{eq-total-facility-power}:
\begin{equation}
\begin{aligned}
    p_{\text{dtc, obj}} = N\cdot (&p_{\text{processor, cp}} +  p_{\text{non-processor, cp}} + p_{\text{sf}}) \ + \\ & p_{\text{cl, air}} + p_{\text{cl, liquid}} + p_{\text{standby}}
\end{aligned}\label{eq-total-facility-power}
\end{equation}

\nomenclature[P]{$p_\text{standby}$}{The power consumption of standby devices in the \acs{DtC}}

\subsubsection{Non-linear model}

\begin{figure}[pos=htbp]
    \centering
    \begin{subfigure}[t]{\linewidth}
        \centering
        \includegraphics[width=.9\linewidth]{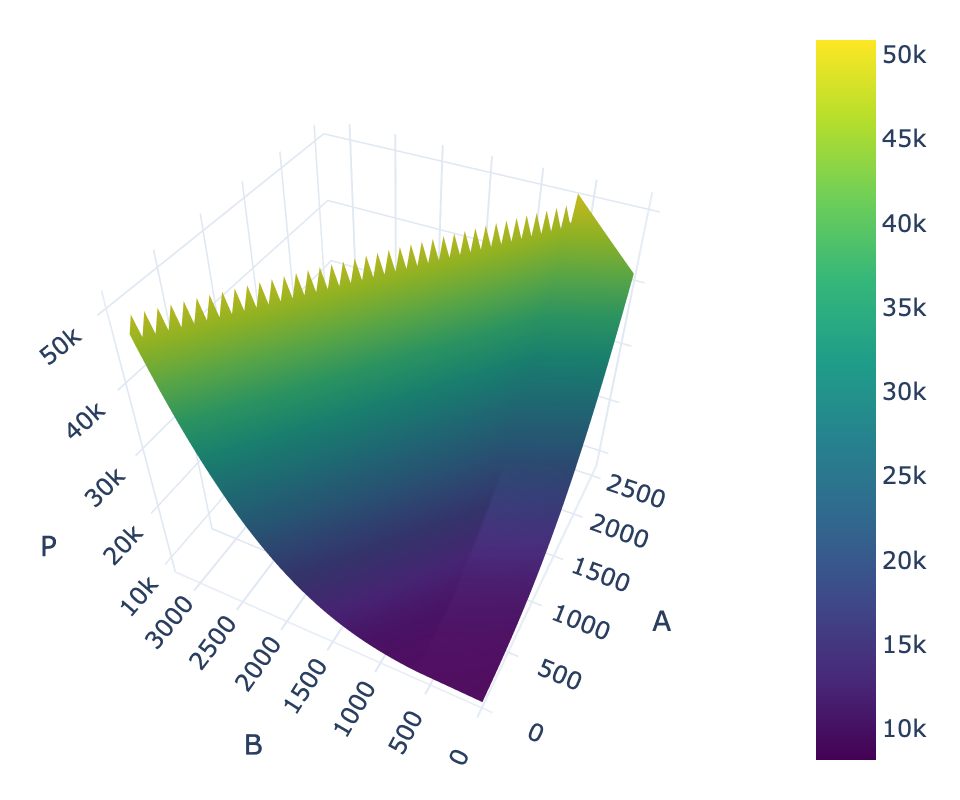}
        \caption{Data center power}\label{fig-p-dtc}
    \end{subfigure}
    
    \begin{subfigure}[t]{\linewidth}
        \centering
        \includegraphics[width=.9\linewidth]{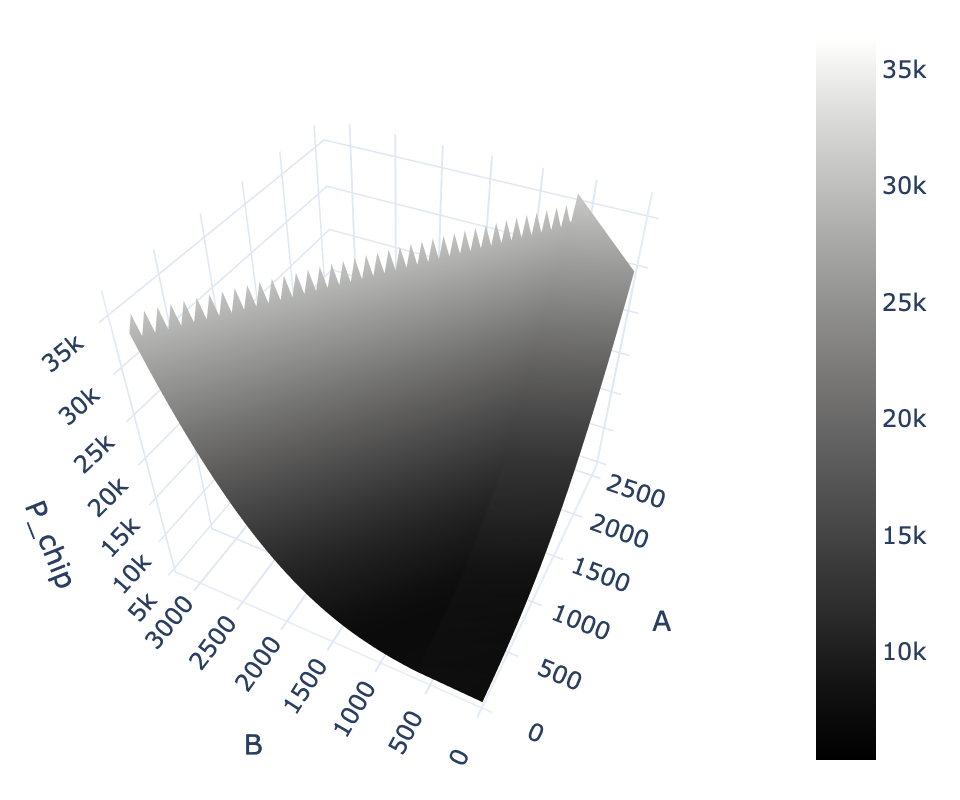}
        \caption{Total processor power (liquid cooled)}\label{fig-p-processor}
    \end{subfigure}
    \caption{An example of result visualization for problem \eqref{eq-MINLP}.}
    \label{fig-MINLP}
\end{figure}

With the built-in optimizer model managing and minimizing the \acs{DtC} internal power consumption, the power of the \acs{DtC} is given by the minimal objective of the \acs{MINLP} problem \eqref{eq-MINLP}:
\begin{subequations}
\label{eq-MINLP}
\begin{align}
\varphi (a,b) = & \min_{T_\text{SA}, N, f, r_{\rm fan}, \dot{m}} \quad  p_{\text{dtc, obj}} \label{eq-MINLP-obj}\\
\textrm{s.t.} \quad & T_\text{SA}^\text{min}\leq T_\text{SA}\leq T_\text{SA}^\text{max} \label{eq-MINLP-TSA}\\
  & N \leq N^{\rm max}  \label{eq-MINLP-N} \\
  & f^\text{min}\leq f\leq f^\text{max} \label{eq-MINLP-f}\\
  & 0\leq r_{\rm fan}\leq r_{\rm fan}^\text{rated} \label{eq-MINLP-pi}\\
  & 0\leq \frac{\lambda}{\mu f}\leq 1 \label{eq-MINLP-u}\\
  & 0\leq t_D \leq D \label{eq-MINLP-tD} \\
  & T_\text{processor} \leq T_\text{processor}^\text{max} \label{eq-MINLP-T-processor}\\
  & T_\text{non-processor} \leq T_\text{non-processor}^\text{max} \label{eq-MINLP-T-non-processor}
\end{align}
\end{subequations}
Constraint \eqref{eq-MINLP-TSA} represents the supply air temperature limits. Constraint \eqref{eq-MINLP-N} limit the requirements for the number of activated servers. Constraints \eqref{eq-MINLP-f} and \eqref{eq-MINLP-pi} represent the lower and upper bounds of the relative processor frequency and fan speed. Constraints \eqref{eq-MINLP-u} and \eqref{eq-MINLP-tD} require that the response time of the interactive workloads should meet \acs{SLA}. \eqref{eq-MINLP-T-processor} and \eqref{eq-MINLP-T-non-processor} limit the temperature of the components. The optimization problem \eqref{eq-MINLP} is solved by commercial solvers by scanning the feasible region of $(a,b)$, then an example $\varphi (a,b)$ is constructed as a surface shown in \Cref{fig-p-dtc}. The corresponding surface of $p_\text{processor, cp}$ is also built for modeling the waste heat recovery, shown in \Cref{fig-p-processor}.

\subsubsection{Linearization} \label{sec-dtc-linearization}

As illustrated in \Cref{fig-p-dtc}, the steady-state power consumption map of the \acs{DtC}, as a non-linear surface, presents challenges for the \acs{RT} optimal control. To facilitate integration into a tractable optimal control framework and improve the computation performance, the power model is linearized using a piecewise approximation method by making reference to \cite{zhouDataCenterLoad2025}. The domain of $(a, b)$ is partitioned into $N_{\text{region}}$ distinct regions $R_i\in\mathcal{R}$. Within each region, the implicit non-linear function $\varphi(a,b)$ is approximated by an explicit linear fit $\hat{\varphi}_i(a,b)$:
\begin{equation}
\varphi(a,b) = \hat{\varphi}_i(a,b) + w_i= \alpha_i a + \beta_i b + \gamma_i + w_i, \ \forall R_i \label{eq-dtc-phi-approx}
\end{equation}
where $\alpha_i$ and $\beta_i$ represent the marginal power increases relative to interactive and batch workloads, respectively, and $\gamma_i$ and $w_i$ are the intercept term and linearization residuals. While increasing $N_{\text{region}}$ would further minimize the linearization error that need to be compensated in the control framework , it would simultaneously increase the computational burden by introducing additional binary variables. 

To formulate the appropriate linear model during optimization, a set of mutually exclusive binary variables $z_i$ is introduced. 
The coupling between the continuous power variable $p_{\text{dtc}}$ and the active region is restricted using the Big-M approach:
\begin{equation}
 \hat{\varphi}_i(a,b) \le p_\text{dtc} \quad \text{for } i=1, \dots, {N_\mathrm{region}} \label{eq-con-dtc-lower}
\end{equation}
\begin{equation}
p_\text{dtc} \le \hat{\varphi}_i(a,b)  +  M\cdot(1 - z_i)  \quad \text{for } i=1, \dots, {N_\mathrm{region}} \label{eq-con-dtc-upper}
\end{equation}
where $M$ is a sufficiently large constant that exceeds the maximum possible power fluctuation of the system. Similarly, the total processor power is built with: 
\begin{equation}
 \hat{\phi}_i(a,b) \le p_\text{dtc, cp} \quad \text{for } i=1, \dots, {N_\mathrm{region}} \label{eq-con-dtc-cp-lower}
\end{equation}
\begin{equation}
p_\text{dtc, cp} \le \hat{\phi}_i(a,b)  +  \bar{M}\cdot(1 - z_i)  \quad \text{for } i=1, \dots, {N_\mathrm{region}} \label{eq-con-dtc-cp-upper}
\end{equation}
where $\phi(a,b)$ is a mapping function for the total processor power, $\hat{\phi}_i(a,b)$ is its linear fit, and $\bar{M}$ a large constant.

\nomenclature[P]{$N_\mathrm{region}$}{The number of regions to partition and linearize the \acs{DtC} power model}
\nomenclature[V]{$p_\text{dtc},\ p_\text{dtc, cp}$}{The variable of the \acs{DtC} total power and computing power}
\nomenclature[P]{$M, \bar{M}$}{Large constants for the Big-M approach}

\subsection{Battery energy storage system}

The \acs{BESS} provides an important flexibility to the system operation, whose sizing is beyong the scope of this paper. In this work, the \acs{BESS} is represented using a discrete-time "bucket model" shown in \eqref{eq-BESS-system}. \eqref{eq-bess-e-def} and \eqref{eq-bess-p-def} define the energy evolution within the battery, which is the integral of the internal power over the discrete time step length $\Delta T$. To model the power flow between the \acs{BESS} and the AC bus, we account for the charging ($\eta_{\text{ch}}$) and discharging ($\eta_{\text{dch}}$) efficiencies in \eqref{eq-bess-e-def}. \eqref{eq-bess-e-bound} and \eqref{eq-bess-p-bound} enforce the constraints of the \acs{BESS} energy and power bounds. A fundamental requirement of the \acs{BESS} model is that the battery cannot simultaneously charge and discharge, and the associated mutual exclusivity is guaranteed by introducing a binary variable $z_\text{bess}$ and the Big-M formulation in \eqref{eq-bess-z-def} to \eqref{eq-bess-p-dch-M}, where $z_{\text{bess}}^t = 1$ indicates the charging state, and $z_{\text{bess}}^t = 0$ indicates the discharging state.
\begin{subequations}
\label{eq-BESS-system}
\begin{align}
& soc^{t+\Delta T}  = soc^t + \frac{\Delta T}{ \mathrm{E}_\text{bess}}\cdot (p_\text{bess,ch}^t\cdot\eta_\text{ch} - p_\text{bess,dch}^t/\eta_\text{dch}) \label{eq-bess-e-def}\\
& p_\text{bess}^t = p_\text{bess,ch}^t - p_\text{bess,dch}^t \label{eq-bess-p-def}\\
& \mathrm{soc}_\text{min}\leq soc^t \leq \mathrm{soc}_\text{max} \label{eq-bess-e-bound}\\
& \mathrm{P}_\text{bess,min}\leq p_\text{bess}^t \leq \mathrm{P}_\text{bess,max} \label{eq-bess-p-bound}\\
& z_\text{bess}^t\in \{0,1\} \label{eq-bess-z-def}\\
& p_\text{bess,ch}^t \leq \mathrm{P}_\text{bess,max}\cdot z_\text{bess}^t \label{eq-bess-p-ch-M}\\
& p_\text{bess,dch}^t \leq -\mathrm{P}_\text{bess,min}\cdot (1-z_\text{bess}^t) \label{eq-bess-p-dch-M}
\end{align}
\end{subequations}
To prevent excessive cycling and reflect the aging of the cells, an operational cost $c_{\text{bess}}^t$ is introduced, which is proportional to the energy processed in each interval:
\begin{equation}
c_{\text{bess}}^t = ({p_{\text{bess,ch}}^t + p_{\text{bess,dch}}^t})/{2} \cdot \Delta T \cdot \pi_{\text{bess}} \label{eq-bess-c}
\end{equation}
where $\pi_{\text{bess}}$ represents the marginal degradation cost\footnote{%
The marginal degradation cost can be estimated from the manufacturer's specifications as
$\pi_{\text{bess}} = C_{\mathrm{rep}} /(2 E_{\mathrm{rated}} N_{\mathrm{cycle}})$,
where $C_{\mathrm{rep}}$ is the replacement cost, $E_{\mathrm{rated}}$ is the rated energy capacity, and $N_{\mathrm{cycle}}$ is the warranted equivalent full-cycle lifetime. This approximation distributes the replacement cost uniformly over the total energy throughput of the battery.}, derived from the replacement cost of the \acs{BESS}.

\nomenclature[V]{$soc$}{The variable of the \acs{BESS} state of charge}
\nomenclature[V]{$\mathrm{E}_\text{bess}$}{The energy capacity of the \acs{BESS}}
\nomenclature[V]{$p_\text{bess,ch}, p_\text{bess,dch}$}{The variables of the \acs{BESS} charging and discharging power}
\nomenclature[V]{$c_{\{\cdot\}}$}{The cost of the unit or division $\{\cdot\}$}
\nomenclature[P]{$\pi_{\{\cdot\}}$}{The unit price of the unit or division $\{\cdot\}$}

\subsection{Waste heat collection and recovery system}

As established in the previous \Cref{sec-liquid-cooling-and-pump-power}, the use of liquid cooling allows for the extraction of medium-grade heat. The total recovered thermal power, $q_{\text{rec}}$, is proportional to the computational power of the \acs{DtC}:
\begin{equation}
    q_\text{rec} = \eta_\text{rec}\cdot p_\text{dtc, cp} \label{eq-heat-rec}
\end{equation}
where $\eta_{\text{rec}}$ represents the heat recovery efficiency. The recovered heat is partitioned into three potential streams:
\begin{itemize}
    \item \acs{ORC} unit ($q_{\text{orc}}$): conversion to electricity via an organic Rankine cycle.
    \item \acs{DH} ($q_{\text{dh}}$): direct thermal supply for the heating demand.
    \item Heat dissipation ($q_{\text{lost}}$): intentional heat rejection when recovery is either technically infeasible or unfavorable.
\end{itemize}
The instantaneous heat balance is given by:
\begin{equation}
q_{\text{rec}} = q_{\text{orc}} + q_{\text{dh}} + q_{\text{lost}} \qquad q_{\text{orc}}, q_{\text{dh}}, q_{\text{lost}} \geq 0 \label{eq-con-heat-balance}
\end{equation}

\nomenclature[V]{$q_{\{\cdot\}}$}{The heat flow to the stream ${\{\cdot\}}$}

\subsubsection{ORC regenerator}
The \acs{ORC} unit is modeled as a multi-modal system characterized by an operational inertia. {Unlike electrochemical storage, an \acs{ORC} requires auxiliary equipment, such as pumps and the working-fluid circulation loop, to reach stable operating conditions before electricity can be generated. }To capture these dynamics while maintaining computational tractability for optimization, we adopt a hierarchical modeling approach \cite{raimondicominesiTwoLayerStochasticModel2018}. This approach separates the system into an upper-level layer for mode selection and a lower-level regulation layer for setpoint tracking.

\paragraph{Upper layer model}

The upper layer operates on a slow time scale with time index $k \in \mathcal{K}$ and determines the discrete operational state of the \acs{ORC} unit, which is modeled as a finite-state automata with four discrete states:
\begin{itemize}
    \item Starting: {a transient phase of duration $t_{\text{start}}$, during which auxiliary equipment is activated and the net electrical output is assumed negligible (we follow the model proposed in \cite{raimondicominesiTwoLayerStochasticModel2018} although it usually consume a small amount of electricity during startup)}.
    \item Running: The steady-state generation phase. The net electrical output $y_{\text{orc}}$ is constrained by the heat supply:
    \begin{equation}
        y_\text{orc}^k \leq \eta_\text{orc}\cdot q_\text{orc}^k \label{eq-orc-y-q-upper}
    \end{equation}
    where $\eta_\text{orc}$ is the efficiency of the unit. The fast dynamics is excluded in the upper layer, which indicates:
    \begin{equation}
        y_\text{orc}^k = u_\text{orc}^k
    \end{equation}
    \item Stopping: a transient phase for shutting down, the net generation is considered negligible.
    \item Stopped: the \acs{ORC} unit is off, and the output is zero.
\end{itemize}
To prevent rapid cycling and hardware degradation, a delay parameter $t_{\text{delay}}$ enforces a minimum operational duration. These logical transitions are tracked via auxiliary integer variables $t_{\mathrm{\scriptscriptstyle ON}}^k$ and $t_{\mathrm{\scriptscriptstyle OFF}}^k$ and a binary switch signal $\delta_{\text{orc}}^k$. The variables are updated according to \eqref{eq-orc-t-on} to \eqref{eq-orc-t-delay}.
    \begin{align} 
    t_{\mathrm{\scriptscriptstyle ON}}^{k+1}=&~\begin{cases} t_{\mathrm{\scriptscriptstyle ON}}^k-1 &\mbox {if} ~\delta_\text{orc}^k=1\\
    t_{\mathrm{ start}} &\mbox {if} ~\delta_\text{orc}^k=0 \end{cases} \label{eq-orc-t-on}\\
    t_{\mathrm{\scriptscriptstyle OFF}}^{k+1}=&~\begin{cases} t_{\mathrm{\scriptscriptstyle OFF}}^k+1 &\mbox {if}~\delta_\text{orc}^k=0\\ 0 &\mbox {if} ~\delta _\text{orc}^k=1. \end{cases} \label{eq-orc-t-off}
    \end{align}
\begin{equation}
\delta_{\text{orc}}^k = 1  \quad \text{if} ~-t_\text{delay}\leq t_{\mathrm{\scriptscriptstyle ON}}^k\leq 0 \label{eq-orc-t-delay}
\end{equation}
And the output power follows \eqref{eq-orc-y-upper}:
\begin{equation}
y_\text{orc} ^k  = 
\begin{cases}
u_\text{orc} ^k, \quad & \text{if}\quad \delta_\text{orc}^k = 1 \land   t_{\mathrm{\scriptscriptstyle ON}}^k\leq 0  ,\\
0, \quad & \text{else}.
\end{cases} \label{eq-orc-y-upper}
\end{equation}
To integrate these logical conditions into the system's optimal control framework, the \acs{ORC} behavior is formulated as a \acs{MLD} system. Using the HYSDEL modeling language \cite{torrisiHYSDELaToolGenerating2004}, the qualitative logic (if-then rules) is transformed into a set of linear inequalities of the form:
\begin{equation}
   \begin{cases}
    x_\text{MLD}^{k+1} = Ax_\text{MLD}^k +  B_u u_\text{MLD}^k +  B_\text{aux} w_\text{MLD}^k + B_\text{aff} \\
    y_\text{MLD}^k = Cx_\text{MLD}^k +  D_u u_\text{MLD}^k +  D_\text{aux} w_\text{MLD}^k +  D_\text{aff}\\
    E_x x_\text{MLD}^k +  E_u u_\text{MLD}^k +  E_\text{aux} w_\text{MLD}^k \leq E_\text{aff}
    \end{cases} \label{eq-MLD}
\end{equation}
where auxiliary variables $w_\text{MLD}^k$ are introduced for converting the logical relations to a set of linear inequalities \cite{bemporadControlSystemsIntegrating1999}.


Operational costs of the \acs{ORC} unit are assumed to be primarily driven by the start-up penalty, $\pi_{\text{orc, st}}$, which captures the wear and energy expenditure of mode transitions:
\begin{equation}
    c_\text{orc}^k = \pi_\text{orc, st}\cdot\max\{\delta_{\text{orc}}^k - \delta_{\text{orc}}^{k-1}, 0\} \label{eq-orc-c}
\end{equation}

\paragraph{Lower layer model}

The lower layer operates on a fast time scale with index $h \in \mathcal{H}$ to track output power setpoints within the \emph{Running} mode. The internal dynamics of the turbine are usually calculated from experimental tests and the system identification. In this work, it is approximated by a first-order transfer function $G(s)$ with gain $K_{\rm orc}$ and time constant $\tau_{\text{orc}}$. After discretization at the sampling rate $\tau_l$, the state-space representation becomes \eqref{eq-sigma-orc-lower}:
\begin{equation}
\Sigma_\text{orc,lower}:
    \begin{cases}
        x_\text{orc}^{h+1} = \Phi\cdot x_\text{orc}^h+  \Gamma\cdot u_\text{orc}^h\\
        y_\text{orc}^h = x_\text{orc}^h
    \end{cases}\label{eq-sigma-orc-lower}
\end{equation}
where $\Phi = \mathrm{exp}(-\tau_l/\tau_{\text{orc}})$, $\Gamma = K_{\rm orc}[1 - \mathrm{exp}({-\tau_l/\tau_{\text{orc}}})]$, and $\mathrm{exp}(\cdot)$ denotes the power of $e$. Similar to \eqref{eq-orc-y-q-upper}, constraint \eqref{eq-orc-y-q-lower} also holds in the lower control layer:
\begin{equation}
    y_\text{orc}^h \leq \eta_\text{orc}\cdot q_\text{orc}^h \label{eq-orc-y-q-lower}
\end{equation}
When the system is in the other modes, the output power $y_\text{orc}$ is considered to be 0:
\begin{equation}
    y_\text{orc}^h = 0 \label{eq-orc-y-0-lower}
\end{equation}

\nomenclature[V]{$u_\text{orc}, y_\text{orc}$}{The power setpoint and output power of the \acs{ORC}}
\nomenclature[V]{$\delta_\text{orc}$}{The binary signal for \acs{ORC} state changes}
\nomenclature[P]{$t_{\mathrm{\scriptscriptstyle ON}}^k, t_{\mathrm{\scriptscriptstyle OFF}}^k$}{The auxiliary integer variables for \acs{ORC} state changes}

\subsubsection{District heating system}
The economic revenue of the recovered heat supplying the \acs{DH} system is modeled as:
\begin{equation}
    c_\text{dh}^t = \pi_\text{dh}^t \cdot q_\text{dh}^t\cdot\Delta T \label{eq-dh-c}
\end{equation}
where $\pi_\text{dh}^t$ represents the time-varying marginal cost of heat production and $\Delta T$ is the discrete time step. This formulation incentivizes the system to increase the heating supply during periods of high heating demand (reflected in high $\pi_\text{dh}^t$ ) and stop the heating supply during non-heating seasons.

\subsection{PV installation}

To balance computational efficiency with physical accuracy in the system-level optimization, the \acs{PV} system is represented using a performance-ratio-based model. The generated power is assumed to scale linearly with the available solar irradiance, as defined by \eqref{eq-pv-model}:
\begin{equation}
    p_\text{pv}^t = \frac{\text{GHI}^t}{\rm{GHI_{ref}}} \mathrm{P}_\text{pv}^\text{rated} \label{eq-pv-model}
\end{equation}
where $\text{GHI}^t$ is the \acs{GHI} at the site in $\unit{W/m^2}$, $\rm{GHI}_{ref}$ is the reference irradiance under standard test conditions and $P_{\text{pv}}^{\text{rated}}$ is the installed peak capacity of the \acs{PV} plant. In scenarios where local generation exceeds the limit of the system, a curtailment term, $p_{\text{pv, curt}}$, represents the intentional curtailment of the generation:
\begin{equation}
    p_\text{pv, out}^t + p_\text{pv, curt}^t = p_\text{pv}^t
\end{equation}
where $p_{\text{pv, out}}$ is the net power injected into the local bus. 

\nomenclature[V]{$p_\text{pv}, p_\text{pv, out}, p_\text{pv, curt}$}{The total available, output, and curtailed power of the \acs{PV}}

\section{Two-layer real-time control framework}\label{sec-two-layer-real-time-control-framework}

\begin{figure}[pos=htbp]
    \centering
    \includegraphics[width=\linewidth]{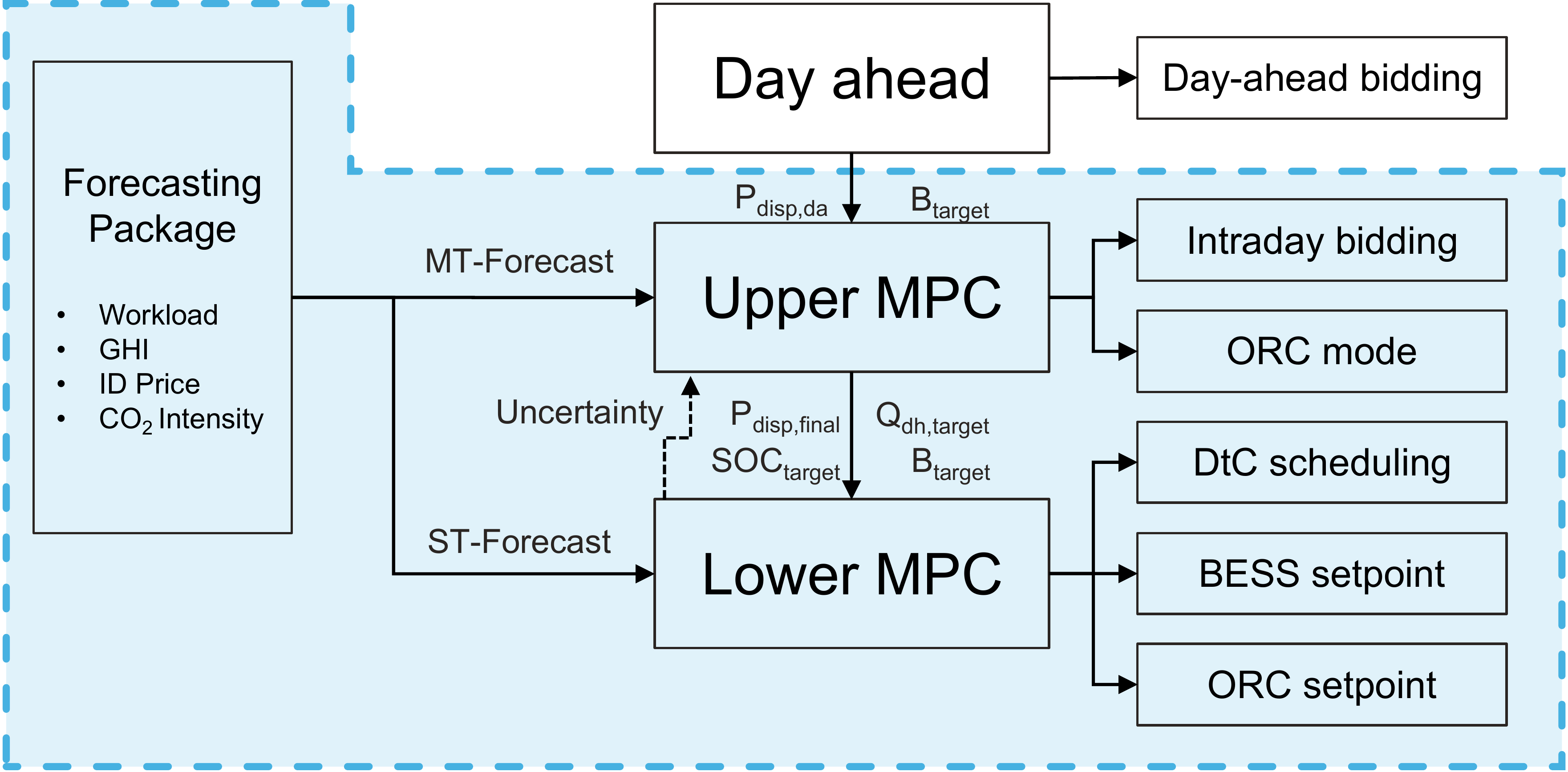}
    \caption{Schematic of the proposed two-layer \acs{RT} control framework.}
    \label{fig-two-layer-control}
\end{figure}

A hierarchical \acs{RT} control framework is proposed to bridge \acs{DA} scheduling and \acs{RT} operation. The framework takes as input a \acs{DA} dispatch plan, including the \acs{PCC} power schedule and booked batch workloads \cite{figiniDayAheadBiddingStrategy2026}, and updates operational decisions in response to forecast uncertainty and \acs{RT} disturbances. As shown in \Cref{fig-two-layer-control}, the upper layer employs scenario-based stochastic \acs{MPC} to optimize dispatch decisions over a longer horizon, while the lower layer uses adaptive tube-based \acs{MPC} to track the resulting references at a higher temporal resolution. This decomposition enables the controller to simultaneously address medium-term (15-min or above) uncertainty and short-term (minute-level) variability while maintaining computational tractability.

\subsection{Upper control layer}

\subsubsection{Problem formulation}

The upper control layer operates with a sampling time of $\tau_u = \qty{15}{min}$ and a shrinking \acs{MPC} horizon spanning the remainder of the day. The discrete time index is denoted by $k = 0, 1, \dots, \bar{k}, \dots, K-1$, where $\bar{k}$ is the current control time step, $K = 96$ corresponds to a full day with 15-minute resolution (also the \acs{MTU} of the \acs{DA} plan) and the stochastic scenario index is denoted by $\omega \in \Omega$. At each execution of the upper layer, the controller uses the most recent \acs{ID} forecasts together with the current system state to compute optimal operational decisions for the next control interval. These decisions include:
\begin{itemize}
\item \acs{ID} market bidding quantities $p_\text{id}^{\bar{k}}$ ,
\item \acs{BESS} SOC targets $soc^{\bar{k}+1}$ ,
\item batch workload execution targets $B_\text{executed}^{\bar{k}+1}$ ,
\item operating state changes of the \acs{ORC} unit $\delta_\text{orc}^{\bar{k}}$,
\item \acs{DH} supply targets $Q_\text{delivered}^{\bar{k}+1}$ .
\end{itemize}

\nomenclature[P]{$\tau_l, \tau_u$}{The sampling time of the lower and upper control layer}
\nomenclature[P]{$h, k$}{The time indices of the lower and upper control layer}
\nomenclature[P]{$\omega, \Omega$}{The stochastic scenarios and the scenario set}

\paragraph{Objective function}
The function of the upper control layer is to track the \acs{DA} dispatch plan while minimizing \acs{ID} operational costs. To account for uncertainties from predictions, a scenario-based stochastic programming framework is adopted. The upper-layer objective function, given in \eqref{eq-upper-objective}, is defined as a weighted combination of the expected \acs{ID} operational cost and the $\alpha$-level \acs{CVaR} over the set of stochastic scenarios $\Omega$:
\begin{equation}
    J_\text{upper} = (1-\beta)\cdot \mathbb{E}\!\left(\Pi\right) +  \beta\cdot\mathrm{CVaR}_\alpha\!\left(\Pi\right)
    \label{eq-upper-objective}
\end{equation}
where $\Pi^\omega$ denotes the total remaining \acs{ID} operational cost associated with scenario $\omega$, $\mathbb{E}(\cdot)$ denotes the expectation operator, and $\beta \in [0,1]$ controls the trade-off between the expected cost optimality and risk aversion.

\nomenclature[P]{$\alpha, \beta$}{Uncertainty tail probability and the weight of \acs{CVaR}}



The \acs{ID} operational cost associated with scenario $\omega$ is defined over the horizon from the current step $\bar{k}$ to the end of the day and is given by \eqref{eq-upper-pi}:
\begin{equation}
\begin{aligned}
        \Pi^\omega & = \Pi^\omega_\text{op} + \Pi^\omega_\text{id} \\
         = &\sum_{k=\bar{k}}^{K-1}  \tau_u\pi_\text{imb}^{k,\omega}\left|p_\text{disp}^{k,\omega} - p_\text{pcc}^{k,\omega}\right| +  \\
         \quad & \sum_{k=\bar{k}}^{K-1} \left[
        c_\text{bess}^{k,\omega} + c_\text{orc}^{k,\omega} - c_\text{dh}^{k,\omega} + c_\text{carbon}^{k,\omega} +  c_\text{b, waste}^{k,\omega}  \right]  + \Pi^\omega_\text{id}
\end{aligned}\label{eq-upper-pi}
\end{equation}
where $\pi_\text{imb}^{k,\omega}$ denotes the imbalance price, $p_\text{disp}^{k,\omega}$ and $p_\text{pcc}^{k,\omega}$ represent the final dispatched and modeled power at \acs{PCC}, respectively, and $c_\text{x}^{k,\omega}$ denotes the cost or revenue associated with component $\text{x}$. The component-specific cost terms are defined in \eqref{eq-bess-c}, \eqref{eq-orc-c}, and \eqref{eq-dh-c}. $c_\text{carbon}^{k,\omega}$ denotes the cost of operational carbon emissions from the upstream grid and is defined in \eqref{eq-c-carbon}:
\begin{equation}
    c_\text{carbon}^{k,\omega} = \tau_u \cdot \pi_\text{carbon} \cdot i_\text{carbon}^k \cdot p_\text{pcc}^{k,\omega} \label{eq-c-carbon}
\end{equation}
where $\pi_\text{carbon}$ represents the carbon price in $\unit{EUR/g CO_2eq}$, $i_\text{carbon}^k$ is the time-variant carbon intensity of the public grid in $\unit{gCO_2eq/kWh}$. $c_\text{b, waste}^{k,\omega}$ is given by \eqref{eq-c-b-waste}, which penalizes the unnecessary batch workload execution since the \acs{DtC} is allowed to execute more batch workload than requested for its flexibility in \eqref{eq-con-b-job-dtc}.
\begin{equation}
    c_\text{b, waste}^{k,\omega} = \tau_u \cdot \pi_\text{b, waste} \cdot (b^{k,\omega} - b^{k,\omega}_\text{job}) \label{eq-c-b-waste}
\end{equation}
In \eqref{eq-c-b-waste}, $\pi_\text{b, waste}$ represents the virtual operational cost of a "computational dump workload", which captures the hidden cost of over-provisioning that incurs wear and tear on hardware and potentially reduces the lifespan of the equipment.
Lastly, $\Pi^\omega_\text{id}$ represents the \acs{ID} market bidding cost for scenario $\omega$ (which is discussed in the following paragraph).

\nomenclature[V]{$p_\text{disp}, p_\text{da}, p_\text{id}$}{The final, \acs{DA}, and \acs{ID} dispatch plan}

\paragraph{Intraday market participation}
\label{sec-intraday-bidding}

The upper-layer controller is allowed to adjust the \acs{DA} dispatch plan by participating in the continuous \acs{ID} electricity market. Ideally, the total remaining \acs{ID} trading cost is given by
\begin{equation}
    \Pi_\text{id}
    =
    \sum_{k=\bar{k}}^{K-1}
    \sum_{\kappa=k+1}^{K-1}
    \pi_\text{id}^{k,\kappa}
    p_\text{id}^{k,\kappa},
    \label{eq-id-cost-ideal}
\end{equation}
where $p_\text{id}^{k,\kappa}$ denotes the transaction volume submitted at time $k$ for delivery during future period $\kappa$, and $\pi_\text{id}^{k,\kappa}$ is the corresponding market price.

Direct optimization over all submission-delivery combinations is computationally intractable for the \acs{RT} \acs{MPC}. Moreover, most transactions in European continuous \acs{ID} markets are executed shortly before delivery \cite{serafinTradingShorttermPath2022,yuOrderbookFeatureLearning2026}. Therefore, in this work, a simplified bidding strategy is adopted in which each delivery interval can be adjusted only once, at a fixed lead time of one hour before delivery. This approximation reduces the number of decision variables while remaining consistent with typical trading practice. For every tradable delivery interval\footnote{The quarter-hourly, half-hourly, and hourly contracts exist at the same time in current European \acs{ID} markets while \acs{DA} markets have switched to a 15-minute \acs{MTU} \cite{epexspot_tradingproducts}. In this work, We only trade hourly \acs{ID} products due to the limited historical transactions in the Swiss market.}, the controller optimizes the net \acs{ID} transaction
\begin{equation}
    p_\text{id}
    =
    p_{\text{id,buy}}
    -
    p_{\text{id,sell}},
\end{equation}
where $p_{\text{id,buy}}\ge0$ and $p_{\text{id,sell}}\ge0$ denote the purchased and sold electricity, respectively. Binary variables are introduced to enforce mutually exclusive buy and sell actions and the minimum bid volume required by the market. The resulting dispatch plan is updated according to
\begin{equation}
    p_\text{disp}^k
    =
    p_\text{da}^k
    +
    p_\text{id}^k,
\end{equation}
subject to the \acs{PCC} import limits
\begin{equation}
    \mathrm{P}_{\text{pcc,min}}
    \le
    p_\text{disp}^k
    \le
    \mathrm{P}_{\text{pcc,max}}.
\end{equation}

The \acs{ID} trading cost under stochastic scenario $\omega$ is then expressed as
\begin{equation}
    \Pi_\text{id}^{\omega}
    =
    \tau_\text{contract}
    \sum_k
    \left(
    p_{\text{id,buy}}^k
    \pi_{\text{id,high}}^{k,\omega}
    -
    p_{\text{id,sell}}^k
    \pi_{\text{id,low}}^{k,\omega}
    \right),
    \label{eq-id-cost}
\end{equation}
where $\pi_{\text{id,high}}$ and $\pi_{\text{id,low}}$ denote the predicted expected buying and selling prices, respectively. Since the study performs a conservative price forecasting (high price for buying and low price for selling, in Appendix \ref{sec-app-forecasting-id-price}), all submitted orders are assumed to be fully matched.

\paragraph{System constraints}

The system operation is constrained by the component models introduced in \Cref{sec-system-model} as well as other limits. The upper layer \acs{MPC} problem is formulated as \eqref{eq-problem-upper}:

\begin{subequations}
\label{eq-problem-upper}
\begin{align}
 \min&  \ (1 - \beta)  \underbrace{\left[\sum_{\omega\in \Omega} \mathrm{p}^\omega \Pi^\omega\right]}_{\mathbb{E}(\Pi)} + \beta \underbrace{\left[ \zeta + \frac{1}{1-\alpha} \sum_{\omega\in \Omega} \mathrm{p}^\omega \theta^\omega \right]}_{\text{CVaR}_\alpha(\Pi)}\\ 
\textrm{s.t.}&  \quad \forall~ k \in \mathcal{K}_{\bar{k}} = \{\bar{k}, \bar{k}+1, ..., K-1 \}, \ \forall~\omega \in \Omega \nonumber \\
& \text{DtC model:}\ \eqref{eq-dtc-wl-dynamics},~\eqref{eq-con-dtc-lower}~\text{to}~\eqref{eq-con-dtc-cp-upper} \label{eq-con-dtc-model-upper}\\
& \text{BESS model:}\ \eqref{eq-BESS-system} \\
& \text{PV model:}\ \eqref{eq-pv-model} \\
& \text{ORC model:}\ \eqref{eq-orc-y-q-upper}~\text{and}~\eqref{eq-MLD}\\
& \text{Heat recovery:}\ \eqref{eq-heat-rec} ~\text{and}~\eqref{eq-con-heat-balance} \label{eq-con-heat-recovery-upper}\\
& \text{Power balance:}\ p_\text{pcc}^{k,\omega} - p_\text{bess}^{k,\omega} + p_\text{pv, out}^{k,\omega} + y_\text{orc}^{k,\omega} = p_\text{dtc}^{k,\omega} \label{eq-con-power-balance-upper}\\
& soc^{K,\omega} = \mathrm{soc}^0  \label{eq-con-soc-terminal-upper}\\
& B_\text{executed}^{\bar{k}, \omega} = \mathrm{B}_\text{executed}^{\bar{k}} \label{eq-con-B-init-upper}\\
& B_\text{executed}^{K, \omega} = \mathrm{B}_\text{target} \label{eq-con-B-terminal-upper}\\
& \text{CVaR ancillary:}\  \theta^\omega\ge 0,~\theta^\omega \ge \Pi^\omega - \zeta \label{eq-con-cvar-ancillary-upper}\\
&\text{Constraint tightening:}\ \eqref{eq-con-upper-p-tightening}~\text{and}~\eqref{eq-con-upper-soc-tightening} \label{eq-con-tightening-upper}\\
& \text{Root node:}\ x^{\bar{k}} = x^{\bar{k}, \omega} \label{eq-con-root-node}
\end{align}
\end{subequations}

Constraints \eqref{eq-con-dtc-model-upper} to \eqref{eq-con-heat-recovery-upper} describe the component-level dynamics, with time indices adapted to the upper layer discretization $k \in \mathcal{K}_{\bar{k}}$ and the scenario $\omega\in\Omega$ is assigned to the decision variables. Constraint \eqref{eq-con-power-balance-upper} enforces system-wide power balance. Constraint \eqref{eq-con-soc-terminal-upper} sets the cyclic energy balance to eliminate opportunistic energy depletion over the optimization horizon, while \eqref{eq-con-B-init-upper} and \eqref{eq-con-B-terminal-upper} enforce the execution requirements of the batch workload. Constraint \eqref{eq-con-cvar-ancillary-upper} introduces auxiliary variables to embed the \acs{CVaR} term into a linear optimization framework, as proposed in \cite{rockafellarConditionalValueatriskGeneral2002}. Constraint \eqref{eq-con-root-node} ensures non-anticipativity by enforcing that control decisions at the current time step are identical across all scenarios ($x$ represents any decision that has the coordinate $k$ and $\omega$).

To make the scenario-based \acs{MPC} problem computationally tractable, a scenario reduction method explained in Appendix \ref{sec-app-stochastic-scenarios} is adopted.

\paragraph{Cross-layer constraint tightening}

Given the hierarchical control structure, the constraint tightening applied in the lower-layer tube \acs{MPC} (see \Cref{sec-constraint-tightening}) must also be reflected in the upper-layer problem. This prevents the upper-layer optimizer from selecting overly optimistic targets that cannot be robustly realized by the lower layer. Prior to solving the upper-layer problem, short-term uncertainties over the next 15-minute interval are estimated. Based on these estimates, worst-case \acs{BESS} power correction bounds are computed for the current upper-layer step:
\begin{equation}
    \Delta\!p_\text{bess, ch}^{\bar{k}}  = \max_{h\in\mathcal{H}} \Delta\!p_\text{bess, ch}^h \quad  \Delta\!p_\text{bess, dch}^{\bar{k}}  = \min_{h\in\mathcal{H}} \Delta\!p_\text{bess, dch}^h
\end{equation}
where $h$ is the discrete time index of the corresponding lower layer control problem, and $\mathcal{H}$ is the set of the indices.
The nominal BESS power limits are then tightened as:
\begin{equation}
\begin{aligned}
    &p_\text{bess, ch}^h\le \mathrm{P}_\text{ch, tight} = \max\left\{\mathrm{P}_\text{bess, max}- \Delta\!p_\text{bess, ch}^{\bar{k}},~0\right\} \\
    & p_\text{bess, dch}^h\le \mathrm{P}_\text{dch, tight} =  \max\left\{-(\mathrm{P}_\text{bess, min}- \Delta\!p_\text{bess, dch}^{\bar{k}}),~0\right\} \label{eq-con-upper-p-tightening}
\end{aligned}
\end{equation}
Uncertainty in BESS power injection propagates directly to its SOC. The resulting worst-case SOC deviations over one upper-layer interval are bounded by:
\begin{equation}
    \begin{aligned}
    & \Delta\!\mathrm{soc}_\text{lb}^{\bar{k}} = \frac{\tau_u}{\eta_\text{dch}\cdot\mathrm{E}_\text{bess}}\cdot \min \left\{ \mathrm{P}_\text{dch, tight},~ \left|\Delta\!p_\text{bess, dch}^{\bar{k}} \right|\right\} \\
    & \Delta\!\mathrm{soc}_\text{ub}^{\bar{k}} = \frac{\eta_\text{ch}\cdot\tau_u}{\mathrm{E}_\text{bess}}\cdot \min \left\{ \mathrm{P}_\text{ch, tight},~ \Delta\!p_\text{bess, ch}^{\bar{k}}\right\}
    \end{aligned}
\end{equation}
The reachable SOC range within the current control interval is therefore:
\begin{equation}
\begin{aligned}
    & \mathrm{soc}_\text{reach,lb}^{\bar{k}} = \mathrm{soc}^{\bar{k}} - \frac{\tau_u}{\eta_\text{dch}\cdot\mathrm{E}_\text{bess}}\cdot\mathrm{P}_\text{dch, tight}\\
    & \mathrm{soc}_\text{reach,ub}^{\bar{k}} = \mathrm{soc}^{\bar{k}} + \frac{\tau_u\cdot\eta_\text{ch}}{\mathrm{E}_\text{bess}}\cdot\mathrm{P}_\text{ch, tight}
\end{aligned}
\end{equation}
Thus, the nominal SOC constraints are tightened for the first control step to ensure robust feasibility in the lower control layer:
\begin{equation}
\begin{aligned}
    soc^{\bar{k}+1} &\le \max\left\{\mathrm{soc}_\text{reach,lb}^{\bar{k}},~ \mathrm{soc}_\text{min} +\Delta\!\mathrm{soc}_\text{lb}^{\bar{k}}\right\}  \\
    soc^{\bar{k}+1} &\ge \min\left\{\mathrm{soc}_\text{reach,ub}^{\bar{k}},~ \mathrm{soc}_\text{max} -\Delta\!\mathrm{soc}_\text{ub}^{\bar{k}} \right\}
\end{aligned}
    \label{eq-con-upper-soc-tightening}
\end{equation}
By using the tightening strategy, the upper-layer SOC targets remain compatible with the physical constraints and tracking capabilities of the lower-layer \acs{MPC}, thereby enhancing cross-layer feasibility and promoting reliable operation.

\subsection{Lower control layer}

The upper-layer controller determines an economically optimal dispatch plan based on medium-term stochastic forecasts. However, forecast errors in \acs{RES} and interactive workload inevitably lead to \acs{RT} deviations from the planned operating point. To compensate for these disturbances while guaranteeing robust operation, the lower control layer employs a tube-based \acs{MPC}.

\subsubsection{Problem formulation}
The lower control layer operates with a sampling time of $\tau_l$ (e.g. $\qty{1}{min}$) and a shrinking \acs{MPC} horizon spanning the remainder of the interval that the corresponding upper layer covers. The discrete time index is denoted by 
\begin{equation}
    h = 0, 1, \dots, \bar{h}, \dots, H-1
\end{equation}
where $\bar{h}$ is the current control step and $H=15$ corresponds to the 15-minute resolution of the upper layer. In contrast to the upper control layer, the lower layer relies on tube-based \acs{MPC}, which is motivated by the fast execution and the availability of \acs{RT} measurements for bounding uncertainties. At each execution, the lower-layer controller computes optimal operational decisions based on the current system state and updated short-term uncertainty bounds. The decision variables include:
\begin{itemize}
    \item \acs{BESS} power setpoints $p_\text{bess}^{\bar{h}}$ ,
    \item batch workload execution setpoints $b^{\bar{h}}$ and $b_\text{job}^{\bar{h}}$ ,
    \item \acs{ORC} power output setpoints $u_\text{orc}^{\bar{h}}$,
    \item \acs{DH} supply setpoints $q_\text{dh}^{\bar{h}}$ .
\end{itemize}

\paragraph{Objective function}

The objective of the lower control layer is to track the power reference at the \acs{PCC} provided by the upper layer, while simultaneously executing the planned batch workload, supplying \acs{DH} demand, and steering the \acs{BESS} SOC toward its target. The overall cost function is given by \eqref{eq-objective-lower}:
\begin{equation}
    J_\text{lower}  =  J_\text{lower, pcc} + J_\text{soc, point}+ J_\text{soc, interval} + J_\text{b, waste} + J_\text{b, flat} \label{eq-objective-lower}
\end{equation}
The individual cost terms are defined as follows:
\begin{itemize}
    \item $J_\text{lower, pcc}$ is given by \eqref{eq-j-lower-pcc}, which penalizes deviations between the actual \acs{PCC} power and the reference dispatched by the upper layer, thereby enforcing tracking:
    \begin{equation}
        J_\text{lower, pcc} =   \tau_l \cdot  \pi_\text{dev}\cdot \left( \sum_{j=\bar{h}}^{H} \left| \mathrm{p}_\text{pcc}^{\bar{k}} - p_\text{pcc}^j\right| \right) \label{eq-j-lower-pcc}
    \end{equation}
    where $\pi_\text{dev}$ is the deviation penalty.
    \item $J_\text{soc, point}$ and $J_\text{soc, interval}$ are given by \eqref{eq-j-soc-point} and \eqref{eq-j-soc-interval}. The two items encourage convergence of the \acs{BESS} SOC to the point-wise target and enforces feasibility with respect to the SOC interval target provided by the upper layer and is discussed in detail in \Cref{sec-constraint-tightening}.
    \item $J_\text{b, waste}$ penalizes unnecessary batch workload execution (note that $b^h \geq b_\text{job}^h$), similar to \eqref{eq-c-b-waste}:
    \begin{equation}
        J_\text{b, waste} = \tau_l \cdot  \pi_\text{b, waste}\cdot \sum_{j=\bar{h}}^{H} \left(b^h - b_\text{job}^h\right) \label{eq-j-b-waste}
    \end{equation}
    \item $J_\text{b, flat}$ promotes a smooth batch workload execution profile. Two auxiliary variables, $b_\text{min}$ and $b_\text{max}$, are introduced such that:
    \begin{equation}
        b_\text{min} \le b^{h} \le b_\text{max} \quad \forall~h\in\mathcal{H}_{\bar{h}} \label{eq-con-b-min-max}
    \end{equation}
    and the corresponding penalty is given by \eqref{eq-j-b-flat}: 
    \begin{equation}
        J_\text{b, flat} = \pi_\text{b, flat} (b_\text{max} - b_\text{min}) \label{eq-j-b-flat}
    \end{equation}
    where $\pi_\text{b, flat}$ is a small virtual cost to be tuned by the modeler.
\end{itemize}

\paragraph{System constraints}

The lower-layer \acs{MPC} problem is formulated as \eqref{eq-problem-lower}:
\begin{subequations}
\label{eq-problem-lower}
\begin{align}
&  \min \ J_\text{lower} \label{eq-problem-lower-obj}\\
\textrm{s.t.} \quad & \forall~ h \in \mathcal{H}_{\bar{h}} = \{\bar{h}, \bar{h}+1, ..., H-1 \} \nonumber \\
& \text{DtC model:}\ \eqref{eq-dtc-wl-dynamics},~\eqref{eq-con-dtc-lower}~\text{to}~\eqref{eq-con-dtc-cp-upper} \label{eq-con-dtc-model-lower}\\
& \text{BESS model:}\ \eqref{eq-BESS-system} \\
& \text{PV model:}\ \eqref{eq-pv-model} \\
& \text{ORC model:}\ \eqref{eq-sigma-orc-lower} ~\text{to}~\eqref{eq-orc-y-0-lower} \\
& \text{Heat recovery:}\ \eqref{eq-heat-rec} ~\text{and}~\eqref{eq-con-heat-balance} \label{eq-con-heat-recovery-lower}\\
& \text{Power balance:}\ p_\text{pcc}^h - p_\text{bess}^h + p_\text{pv, out}^h + y_\text{orc}^h = p_\text{dtc}^h \label{eq-con-power-balance-lower}\\
& soc^{\bar{h}} = \mathrm{soc}^{\bar{h}} \label{eq-con-soc-init-lower}\\
& B_\text{executed}^{\bar{h}} = \mathrm{B}_\text{executed}^{\bar{h}} \label{eq-con-B-init-lower}\\
& B_\text{executed}^{H} = \mathrm{B}_\text{executed}^{\bar{k}+1} \label{eq-con-B-terminal-lower}\\
& Q_\text{delivered}^{\bar{h}} = \mathrm{Q}_\text{delivered}^{\bar{h}} \label{eq-con-q-dh-init-lower}\\
& Q_\text{delivered}^{H} = \mathrm{Q}_\text{delivered}^{\bar{k}+1} \label{eq-con-q-dh-ternimal-lower}\\
& Q_\text{delivered}^{h+1} = Q_\text{delivered}^{h} + \tau_l \cdot q_\text{dh}^h \label{eq-con-q-dh-dynamics-lower}\\
& \text{Ancillary:}\ \eqref{eq-con-b-min-max} \nonumber \\
&\text{Constraint tightening:}\ \eqref{eq-con-lower-p-tightening}~\text{and}~\eqref{eq-con-soc-interval} \label{eq-con-tightening-lower}
\end{align}
\end{subequations}
Constraints \eqref{eq-con-dtc-model-lower} to \eqref{eq-con-heat-recovery-lower} correspond to the component models introduced in \Cref{sec-system-model}, with time indices adapted to the lower layer indexing system $h\in\mathcal{H}_{\bar{h}}$. Constraint \eqref{eq-con-power-balance-lower} ensures the power balance of the system. Constraint \eqref{eq-con-soc-init-lower} sets the initial state of the \acs{BESS} using the measured data. Constraints \eqref{eq-con-B-init-lower} and \eqref{eq-con-B-terminal-lower} enforce the batch workload execution plan provided by the upper layer, while \eqref{eq-con-q-dh-init-lower} to \eqref{eq-con-q-dh-dynamics-lower} ensure consistency with the district heating schedule from the upper layer. Finally, the constraint tightening terms in \eqref{eq-con-tightening-lower} account for bounded uncertainties under the tube-based \acs{MPC} framework and are discussed in detail in \Cref{sec-constraint-tightening}.

\subsubsection{Adaptive tube MPC for the lower control layer}

\nomenclature[P]{$K$}{The feedback gain of the tube \acs{MPC} local feedback}

The key idea of tube \acs{MPC} is to decouple nominal trajectory planning from disturbance rejection. The optimization problem first predicts a nominal system trajectory assuming no disturbances, while a local feedback controller continuously compensates deviations caused by model mismatch and forecast errors. Under bounded disturbances, the feedback correction and corresponding constraint tightening keep the actual trajectory within a neighborhood of the nominal trajectory \cite{langsonRobustModelPredictive2004}. Accordingly, the implemented control input consists of a nominal control action and a local feedback policy,
\begin{equation}
    \tilde{u}^h = u^h + K e^h,
\end{equation}
where $u^h$ denotes the nominal control input obtained from the \acs{MPC} optimization, $e^h$ is the deviation between the actual and nominal trajectories, and $K$ is a feedback gain designed offline. The nominal optimization is solved over tightened state and input constraints, while the feedback controller compensates \acs{RT} disturbances and keeps the actual trajectory within the disturbance tube. { In the proposed framework, $K$ is kept fixed, whereas the disturbance set is adapted online. The two quantities serve different purposes: $K$ determines the response of the local feedback policy, while the disturbance set determines the magnitude of the correction that must be reserved by the nominal controller. A fixed gain therefore allows the local feedback policy to remain unchanged while the error margin adapts to the latest forecast uncertainty.

In this study, the \acs{BESS} is the main leveraged assest due to its fast actuation capability to compensate short-term power imbalance and the disturbances affecting the lower-layer controller are represented as an equivalent uncertainty in the power balance at the \acs{PCC}. Therefore, the gain $K$ is selected offline according to the available \acs{BESS} correction capability and the expected disturbance magnitude\footnote{In general, the value of $K$ ($0<K<1$) can be selected using standard feedback design methods accounting for actuator limits \cite{langsonRobustModelPredictive2004}, disturbance magnitudes, and the desired disturbance-rejection characteristics. In this work, $K=0.99$ is adopted based on the available \acs{BESS} capability relative to the historically estimated disturbances, as well as the willingness of the \acs{DtC} operator to reserve battery capacity for \acs{RT} compensation. A larger $K$ provides stronger disturbance rejection but reserves more \acs{BESS} power for \acs{RT} compensation \eqref{eq:bess-reserved-power}.}.} The actual \acs{PV} generation and \acs{DtC} power consumption are modeled as
\begin{equation}
\begin{aligned}
    \tilde p_{\mathrm{pv}}^h &= p_{\mathrm{pv}}^h + w_{\mathrm{pv}}^h,\\
    \tilde p_{\mathrm{dtc}}^h &= p_{\mathrm{dtc}}^h + w_{\mathrm{dtc}}^h,
\end{aligned}
\end{equation}
where $p_{\mathrm{pv}}^h$ and $p_{\mathrm{dtc}}^h$ denote the nominal predictions obtained from the lower-layer optimization, while $w_{\mathrm{pv}}^h$ and $w_{\mathrm{dtc}}^h$ are forecast error bounds. The aggregate disturbance acting on the \acs{BESS} is therefore
\begin{equation}
    w^h = w_{\mathrm{pv}}^h - w_{\mathrm{dtc}}^h
    \in
    \mathcal W^h
    =
    \mathcal W_{\mathrm{pv}}^h
    \oplus
    (-\mathcal W_{\mathrm{dtc}}^h),
    \label{eq-tube-w-agg}
\end{equation}
which directly determines the correction required to maintain the scheduled power exchange at the \acs{PCC}. The implemented battery power follows the feedback law
\begin{equation}
    \tilde p_{\mathrm{bess}}^h
    =
    p_{\mathrm{bess}}^h
    +
    K
    p_{\mathrm{pcc,error}}^h,
\end{equation}
where $p_{\mathrm{bess}}^h$ is the nominal control action provided by the lower-layer \acs{MPC}, while $p_{\mathrm{pcc,error}}^h$ is computed online from measured system powers.

\paragraph{Online uncertainty estimation}

Unlike conventional tube \acs{MPC}, where the disturbance set is assumed to be fixed, the proposed framework updates the disturbance tube (disturbance sets $\mathcal W_{\mathrm{pv}}^h$ and $\mathcal W_{\mathrm{dtc}}^h$) online according to the latest short-term forecasts. Consequently, the degree of conservatism automatically adapts to the current operating conditions, improving dispatch tracking while avoiding unnecessary constraint tightening during periods of low uncertainty. 

For the \acs{PV} generation, the uncertainty is obtained from the predicted \acs{GHI} interval Appendix \ref{sec-app-forecasting-short-ghi}. Neglecting curtailment, the power uncertainty is approximated through a first-order sensitivity model,
\begin{equation}
    w_{\mathrm{pv}}^h
    \approx
    s_{\mathrm{pv}}
    e_{\mathrm{ghi}}^h,
\end{equation}
where $e_{\mathrm{ghi}}^h$ denotes the \acs{GHI} forecast error and
\begin{equation}
    s_{\mathrm{pv}}
    =
    \frac{\partial p_{\mathrm{pv}}}
    {\partial \mathrm{GHI}}
    =
    \frac{P_{\mathrm{pv}}^{\mathrm{rated}}}
    {\mathrm{GHI}_{\mathrm{ref}}}.
\end{equation}

Similarly, the uncertainty of the \acs{DtC} power consumption is obtained by propagating the forecast interval of the interactive workload through the piecewise-linear power model,
\begin{equation}
    w_{\mathrm{dtc,wl}}^h
    \approx
    s_{\mathrm{dtc}}
    e_{\mathrm a}^h,
\end{equation}
where $e_{\mathrm a}^h$ denotes the workload forecasting error bounds and $s_{\mathrm{dtc}}$ is conservatively selected as the maximum slope among all linearization regions in \Cref{tab-linearization}. The linearization error of the \acs{DtC} model is incorporated as an additional bounded disturbance, yielding
\begin{equation}
    \mathcal W_{\mathrm{dtc}}^h
    =
    \mathcal W_{\mathrm{dtc,wl}}^h
    \oplus
    \mathcal W_{\mathrm{dtc,lin}},
\end{equation}
{where $\mathcal W_{\mathrm{dtc,lin}} = [\min_i w^i_{\rm lb}, \max_i w^i_{\rm ub}]$ (see in \Cref{tab-linearization}, Appendix \ref{sec-app-linearization}).}

Finally, the adaptive disturbance tube is obtained from $\mathcal W^h = \left[ \underline w^h,\overline w^h\right]$, which is computed at each lower control step.

\paragraph{Adaptive constraint tightening}  \label{sec-constraint-tightening}

The estimated disturbance set is translated into admissible \acs{BESS} power corrections,
\begin{equation}
\label{eq:bess-reserved-power}
    \Delta p_{\mathrm{ch}}^h
    =
    K\overline w^h,
    \qquad
    \Delta p_{\mathrm{dch}}^h
    =
    K\underline w^h,
\end{equation}
which are used to tighten the nominal charging and discharging limits,
\begin{equation}
\begin{aligned}
p_{\mathrm{bess,ch}}^h &\le \max \left\{0, P_{\mathrm{bess,max}} - \Delta p_{\mathrm{ch}}^h\right\},\\
p_{\mathrm{bess,dch}}^h &\le \max \left\{0,  |P_{\mathrm{bess,min}}| - |\Delta p_{\mathrm{dch}}^h|\right\}.
\end{aligned} \label{eq-con-lower-p-tightening}
\end{equation}
{\color{black}Thus, for disturbances contained in the estimated set, the nominal optimization reserves sufficient \acs{BESS} power headroom for the local feedback correction. Because the disturbance bounds vary with the forecast horizon, enforcing a hard terminal SOC equality at the lower layer would unnecessarily restrict the feasible set. Instead, the controller adopts a hierarchical terminal objective supplied by the upper layer. }The upper controller provides an ideal terminal SOC together with an admissible interval reflecting the estimated uncertainty. Deviations outside the interval are penalized more strongly, while deviations from the ideal point target receive a smaller penalty:
\begin{equation}
    J_\text{soc, point} = \pi_\text{bess, vio-point} \cdot \left|\mathrm{soc}^H - \mathrm{soc}_\text{target}^{\bar{k}} \right| \label{eq-j-soc-point}
\end{equation}
where $\pi_\text{bess, vio-point}$ is the violation cost of the point target. The interval SOC target is formulated by introducing slack variables $\varepsilon_+$ and $\varepsilon_-$ in \eqref{eq-j-soc-interval} and \eqref{eq-con-soc-interval}:
\begin{equation}
    J_\text{soc, interval} = \pi_\text{soc, vio-interval} \cdot \left(\varepsilon^-+\varepsilon^+\right) \label{eq-j-soc-interval}
\end{equation}
\begin{equation}
    \begin{cases}
        \varepsilon_+\geq 0,\  \mathrm{soc}^H \leq \mathrm{soc}_\text{target, ub}^{\bar{k}} + \varepsilon_+\\
         ~\varepsilon_- \geq 0,\  \mathrm{soc}^H \geq \mathrm{soc}_\text{target, lb}^{\bar{k}} - \varepsilon_-
    \end{cases}\label{eq-con-soc-interval}
\end{equation}
where $\pi_\text{bess, vio-point} <\!< \pi_\text{bess, vio-interval}$. 

The adaptive constraint tightening is recomputed at each lower-layer control step, and the corresponding feasible region may expand or contract as the uncertainty estimates change. In the adopted short-term forecasting framework (Appendix~\ref{sec-app-forecasting}), uncertainty bounds generally widen with the prediction horizon, forming an uncertainty cone that anticipates larger deviations at less recent steps. As the horizon advances, the uncertainty associated with a given future instant typically decreases, although nestedness between successive estimates is not strictly guaranteed. Moreover, the lower layer optimization adopts a soft tracking formulation, as commonly used in practical \acs{MPC} applications \cite{zeilingerSoftConstrainedModel2014,rawlings2017model}: deviations from the scheduled \acs{PCC} exchange are penalized in the objective, while terminal SOC targets are softened using slack variables. Consequently, when increased uncertainty restricts the available nominal \acs{BESS} power under extreme conditions, the controller can reduce its dispatch, including to zero, and tolerate a larger tracking deviation at \acs{PCC} rather than enforcing an infeasible reference. For feasible tightened constraints and feedback corrections within the reserved bounds, \eqref{eq:bess-reserved-power}-\eqref{eq-con-lower-p-tightening} ensure compliance with the physical \acs{BESS} power limits. From the following simulation studies, these mechanisms support operational feasibility under the estimated disturbance range, while a formal recursive feasibility guarantee under arbitrary uncertainty updates is not claimed in this work.

\section{Simulation study}

\subsection{Microservice-based simulation environment}

\subsubsection{Testing framework}
To evaluate the performance of the proposed \acs{RT} control framework, we developed a modular, microservice-based testing environment. This architecture decouples the control logic from the system dynamics and the environmental data sources, ensuring a high-fidelity representation of a real-world deployment where the controller must interact with external systems via standard communication protocols. Moreover, by hosting the simulator as an independent service, a clean boundary between the models used for optimization (which are simplified for convexity and computational speed) and the emulation models (which include non-linearities and higher-order dynamics) is maintained, which provides a rigorous test for the controller's robustness against model plant mismatch. For example, the designed simulator consists of the following component models:
\begin{itemize}
\item \acs{DtC}: the workload execution dynamics follow \eqref{eq-dtc-wl-dynamics}. The electricity consumption ~$\varphi$ is obtained by solving the nonlinear power model \eqref{eq-MINLP} in each simulation step.
\item \acs{PV}: the \acs{PV} power output is modeled using \eqref{eq-pv-model}, where~$\mathrm{GHI}^t$ is provided by real-world measurements with a resolution of 1 minute.
\item \acs{BESS}: an \acs{ECM} with internal resistance, as proposed in \cite{reyes-chamorroComposableMethodRealtime2015}, is used to describe the \acs{BESS} power behavior and SOC dynamics.
\item \acs{ORC}: the \acs{ORC} operating modes are updated according to the logical relations defined in \eqref{eq-orc-t-on} to \eqref{eq-orc-t-delay}, while the electrical output power follows the first-order dynamics given in \eqref{eq-sigma-orc-lower}.
\end{itemize}

\begin{figure}[pos=ht]
    \centering
    \includegraphics[width=\linewidth]{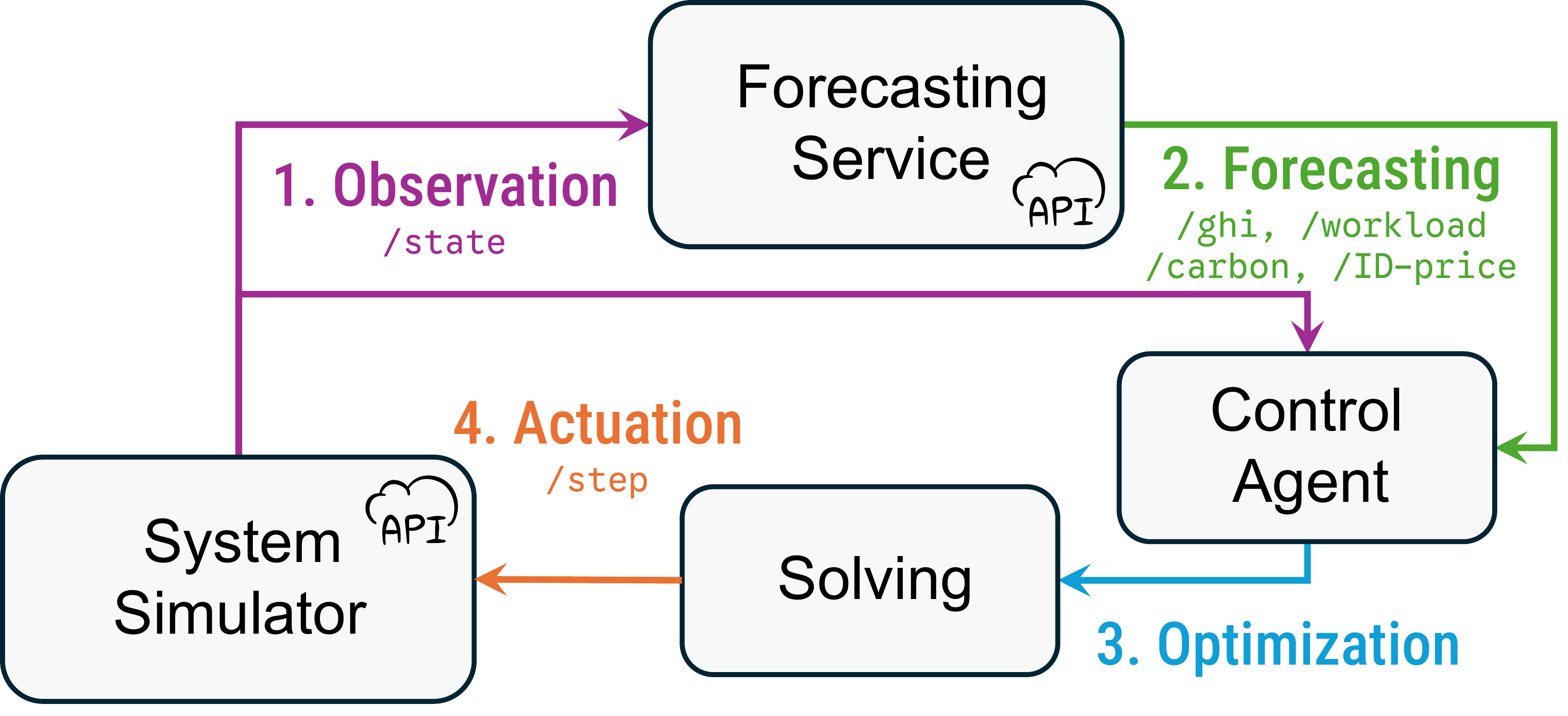}
    \caption{Workflow of the testing framework.}
    \label{fig-execution-cycle}
\end{figure}

The testing framework is composed of three primary entities: the control agent, the forecasting service, and the system simulator. Both the forecasting and simulation components are implemented as standalone applications that expose their functionalities through RESTful interfaces \cite{fieldingArchitecturalStylesDesign2000}. As illustrated in \Cref{fig-execution-cycle}, the communication flow follows an execution cycle:
\begin{enumerate}[i)]
    \item Observation: At the start of each control interval, the controller queries the simulator’s \verb|/state| endpoint to retrieve the current state of the \acs{BESS}, \acs{ORC}, and \acs{DtC}.
    \item Forecasting: Simultaneously, the controller fetches the latest predictions (e.g., carbon intensity, \acs{GHI}, and \acs{ID} electricity prices) from the forecasting API.
    \item Optimization: Using the gathered data, the controller formulates and solves the \acs{MPC} optimization problem described in \Cref{sec-two-layer-real-time-control-framework}.
    \item Actuation: The resulting setpoints are sent to the simulator’s \verb|/step| endpoint, which advances the plant state based on higher-fidelity models to avoid inverse crime.
\end{enumerate}
{ In addition, a hierarchical fallback mechanism is designed to handle communication, solver, or hardware failures. If the lower-layer optimization fails, the corresponding upper-layer dispatch is used as the fallback reference. If the upper-layer optimization fails, the latest feasible upper-layer solution is retained and propagated to the current control step. Although this solution may produce infeasibility since the operating conditions have changed. The optimization is then retried at the subsequent \acs{MPC} step. This mechanism prevents transient computational or communication failures from directly interrupting operation.}

\subsubsection{Data and forecasting}

\begin{table}[ht]
\centering
\caption{Summary of the testing setup.}
\label{tab-simulation_setup}
\footnotesize
\begin{tabular}{lll}
\toprule
\textbf{Category} & \textbf{Parameter} & \textbf{Description / Source} \\
\midrule

\multirow{4}{*}{\acs{DtC}}
& Workload & Synthetic based on \cite{clusterdata:Wilkes2020a} \\
& Server & SPEC test \cite{specSPECPowerHuawei}, $N=32$ \\
& Cooling design &  Confidential \\
& Max power & \qty{\sim 50}{kW} \\

\midrule

\multirow{3}{*}{\acs{PV}}
& Coordinate & $(\qty{46.5184}{\degree N}, \qty{6.5652}{\degree E})$ \\
& Rated P & \qty{50}{kWp} \\
& Ground truth & Measured GHI \\

\midrule

\multirow{3}{*}{\acs{BESS}}
& Size & \qty{40}{kW} / \qty{80}{kWh}\\
& \acs{ECM} & Confidential \\
& $\pi_{\rm bess}$ & \qty{0.002}{EUR/kWh} \\

\midrule

\multirow{4}{*}{\acs{ORC}}
& Output power & $\qty{1.0}{kW}$ to $\qty{3.5}{kW}$ \\
& Efficiency & $\qty{10}{\percent}$\\
& Upper model & $t_{\rm delay} =0$, $t_{\rm start} =1$\\
& Lower model & $\tau_{\rm orc} = \qty{5}{min}$ \\

\midrule

\multirow{2}{*}{\acs{PCC}}
& Max import & \qty{100}{kW} \\
& Max export & \qty{0}{kW} \\

\midrule

\multirow{6}{*}{Market}
& \acs{DA} price & \cite{entsoeTransparencyPlatform}, bidding zone CH \\
& \acs{ID} price & \cite{energy-chartsEnergyChartsAPI}, bidding zone CH \\
& $i_\text{carbon}$ & \cite{electricitymaps2025data}, zone CH \\
& \acs{DH} price & $\qty{0.03}{EUR/kWh}$ \cite{TarifeFernwaerme} \\
& Imb. price &  Fixed $\qty{0.5}{EUR/kWh}$\\
& Carbon price & Baseline $\qty{265}{EUR/tCO_2eq}$ \cite{rennertComprehensiveEvidenceImplies2022} \\
\midrule

\multirow{5}{*}{\acs{MPC}}
& CVaR & Tail $\alpha=0.1$, weight $\beta=0.25$ \\
& $\pi_\text{dev}$ & $\qty{0.5}{EUR/kWh}$ \\
& $\pi_\text{b, waste(flat)}$ & $\qty{4(1)e-3}{EUR/unit}$ \\
& $\pi_\text{soc, vio-interval(point)}$ & $\qty{80(8)e-3}{EUR/kWh}$ \\
& $N_{\rm reduced}$ & 20 (from Appendix \ref{sec-app-scenario-number-sensitivity})\\

\bottomrule
\end{tabular}
\end{table}

The configuration of the simulated system is summarized in \Cref{tab-simulation_setup}\footnote{In \Cref{tab-simulation_setup}, the selected component ratings correspond to the laboratory-scale demonstration platform developed within the Heating Bits project \cite{HeatingBits}. Although the installed capacities are smaller than those of commercial hyperscale data centers, the proposed control framework is independent of the absolute system size. The study therefore focuses on validating the hierarchical control strategy and its interaction with uncertainty rather than assessing the economic viability of market participation for a specific installation.}, while the parameters of the piecewise linearized \acs{DtC} power model are reported in Appendix \ref{sec-app-linearization}. The \acs{DtC} workload traces are derived from the Google Cluster Trace (May 2019, cell H) \cite{clusterdata:Wilkes2020a}. The original 5-min measurements are interpolated to a 1-min resolution and remapped into interactive (Google's \emph{monitoring} and \emph{mid}) and batch (Google's \emph{best-effort} and \emph{free}) workloads. To construct a representative simulation dataset for 2026, typical weekday profiles are aligned with the 2026 calendar using a calendar-matching heuristic, after which independent white Gaussian noise is added to the interpolated data for emulating realistic short-term variability. Moreover, \acs{DA} dispatch plans of the \acs{PCC} power are computed using the method proposed in \cite{figiniShouldSmallScaleData2026}. 

The proposed controller relies on forecasts of four exogenous variables: (i) interactive workload arrivals, (ii) \acs{GHI} for \acs{PV} generation estimation, (iii) \acs{ID} electricity prices, and (iv) grid carbon intensity. Batch workload requests are assumed to be determined by the upstream \acs{DA} scheduling stage and are therefore treated as known inputs. The forecasting methods adopted in this study are described in Appendix \ref{sec-app-forecasting}.



\subsection{Day I: Clear-sky operating conditions } \label{sec-day-I}

\begin{figure}[pos=ht]
    \centering
    \includegraphics[width=\linewidth]{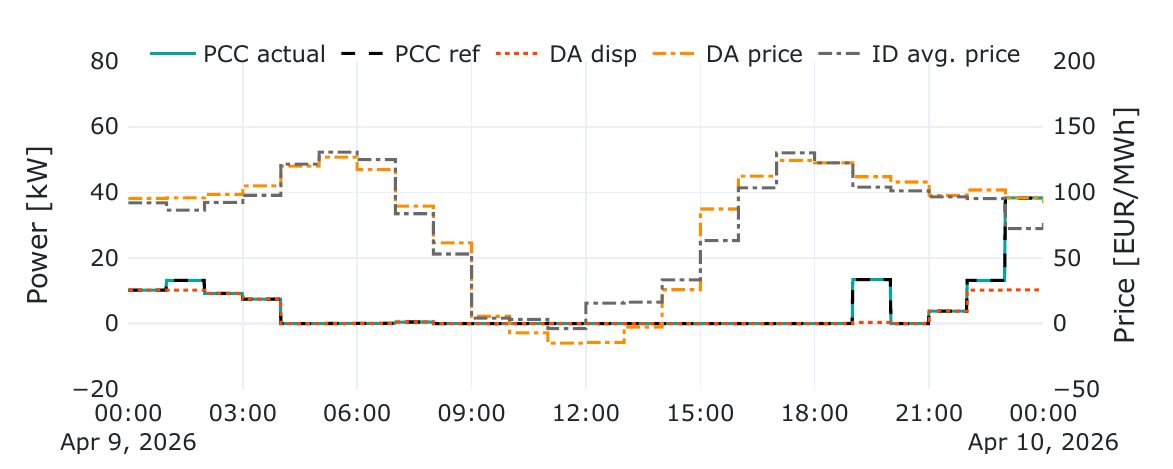}
    \caption{\acs{DA} plan, \acs{ID}-adjusted dispatch reference, realized \acs{PCC} power, and market prices of Apr.9, 2026.}
    \label{fig-pcc-da-disp-2026-04-09}
\end{figure}

\begin{figure}[pos=ht]
    \centering
    \includegraphics[width=\linewidth]{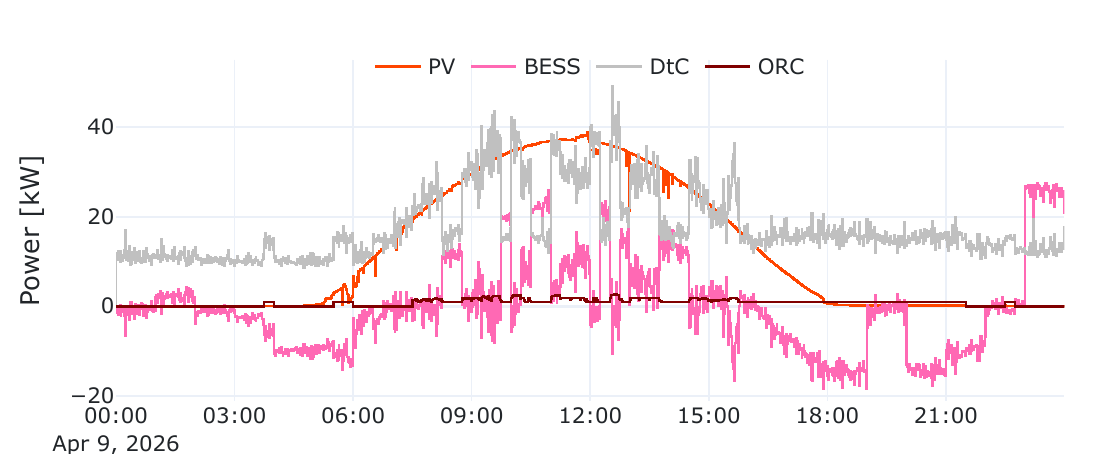}
    \caption{System operation over the day of Apr.9, 2026.}
    \label{fig-operation_2026-04-09_two_layer_stochastic}
\end{figure}

April 9th, 2026 is selected as a representative clear-sky operating day to evaluate the proposed framework under favorable on-site \acs{PV} generation conditions. The system operation over the entire day is illustrated in \Cref{fig-pcc-da-disp-2026-04-09} and \Cref{fig-operation_2026-04-09_two_layer_stochastic}. The controller coordinates the \acs{PV} generation, \acs{BESS}, flexible workloads, and waste heat recovery to maximize local renewable utilization while tracking the scheduled power exchange at the \acs{PCC}. The \acs{BESS} acts as the primary balancing resource by charging during periods of abundant solar generation and discharging when local generation is insufficient. Meanwhile, batch workloads are preferentially shifted to daylight hours to increase on-site consumption of solar energy. Since the system operates outside the heating season (no \acs{DH} demamd), part of the recovered waste heat is converted into electricity by the \acs{ORC} turbine. Although the contribution of the \acs{ORC} remains relatively small due to its limited conversion efficiency, it further reduces the dependence on external grid electricity.


The interaction with the electricity market is also shown in \Cref{fig-pcc-da-disp-2026-04-09}, which compares the \acs{DA} dispatch plan, the \acs{ID}-adjusted dispatch reference, and the realized power exchange at the \acs{PCC}. In addition, the \acs{DA} market clearing price and the average \acs{ID} transaction price reported by ENTSO-E and EPEX Spot \cite{entsoeTransparencyPlatform, energy-chartsEnergyChartsAPI} are shown. The proposed framework closely tracks the dispatch reference throughout the day despite uncertainties in \acs{PV} output and interactive workloads. The largest \acs{ID} adjustment occurs during the final operating hour, where relatively low \acs{ID} prices are exploited to recover the battery SOC to its initial value for the following day. This demonstrates the ability of the hierarchical controller to jointly coordinate market participation and \acs{RT} operation.


\paragraph{Case study: benefits of the two-layer structure}

\begin{figure}[pos=ht]
    \centering
    \includegraphics[width=\linewidth]{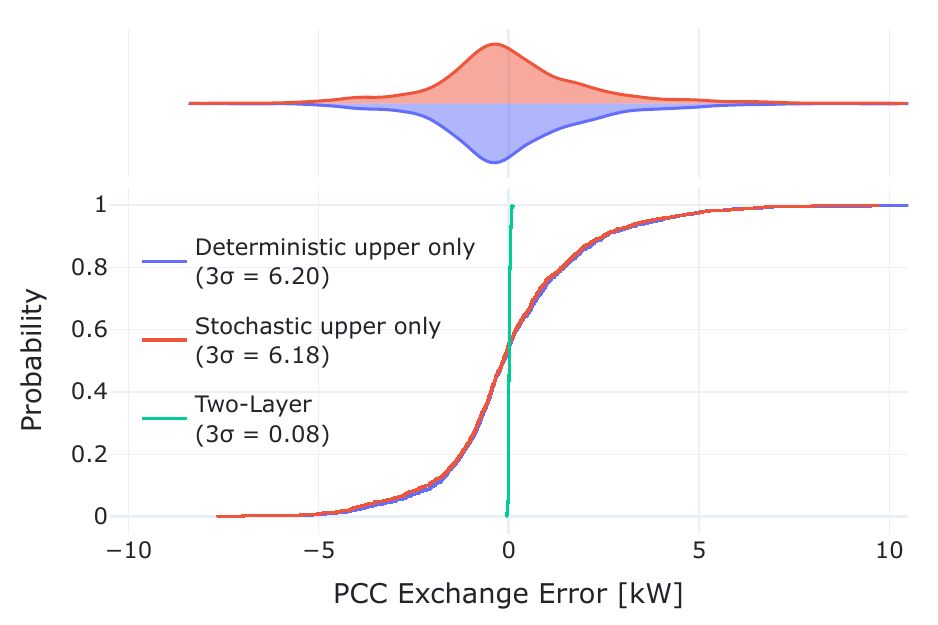}
    \caption{\acs{ECDF} plot of power tracking errors at \acs{PCC}.}
    \label{fig-pcc-error-ecdf-2026-04-09}
\end{figure}

The benefit of the hierarchical control architecture is evaluated by comparing three control strategies: a deterministic upper-layer \acs{MPC}, a stochastic upper-layer \acs{MPC}, and the proposed two-layer framework. \Cref{fig-pcc-error-ecdf-2026-04-09} presents the \acs{ECDF} of the \acs{PCC} tracking error. The proposed framework consistently achieves smaller tracking errors across the entire distribution. Quantitatively, the $3\sigma$ tracking error is reduced from \qty{6.20}{kW} and \qty{6.18}{kW} for the deterministic and stochastic upper-layer controllers, respectively, to only \qty{0.08}{kW} with the proposed two-layer framework. Considering the \acs{DtC} peak power of approximately \qty{50}{kW}, this corresponds to reducing the tracking error from about \qty{12}{\percent} of the peak \acs{DtC} demand to below \qty{0.2}{\percent}. This improvement is achieved by the adaptive tube-based lower controller, which continuously compensates short-term disturbances using the fast response capability of the \acs{BESS}. In contrast, the upper-layer-only controllers rely solely on nominal dispatch references and therefore cannot effectively compensate fast forecast errors and model mismatches that arise between consecutive upper-layer optimization updates.

\begin{figure}[pos=ht]
    \centering
    \includegraphics[width=\linewidth]{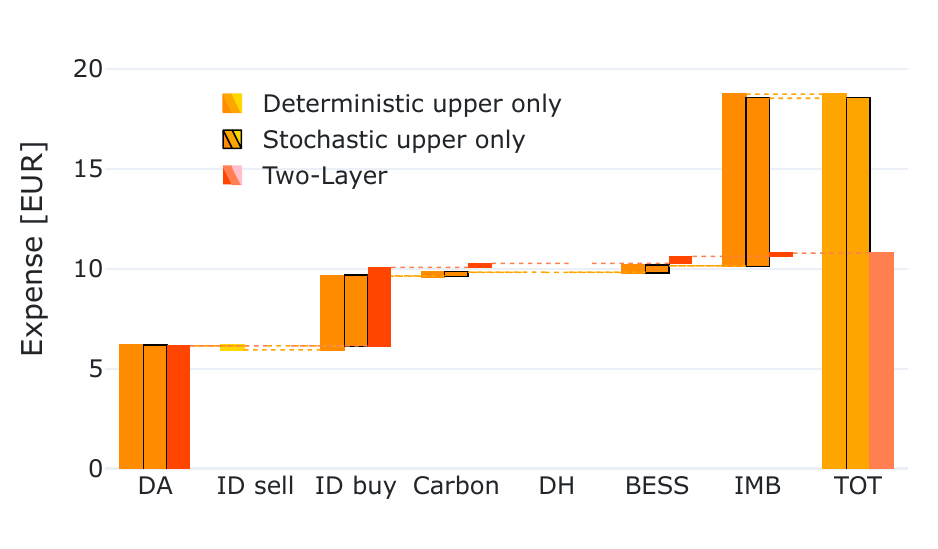}
    \caption{Ex post operational expenses breakdown of Apr.9, 2026.}
    \label{fig-expense_breakdown_2026-04-09_two_layer_stochastic}
\end{figure}

The resulting operational expenses are compared in \Cref{fig-expense_breakdown_2026-04-09_two_layer_stochastic}, where a fixed imbalance tariff of \qty{0.5}{EUR/kWh} is applied to the average energy deviation within each \acs{MTU}. While the three controllers exhibit similar costs associated with \acs{DA} bidding, \acs{ID} trading, \acs{BESS} degradation, and carbon emissions, the imbalance penalty differs significantly. The proposed two-layer framework substantially reduces imbalance costs by improving \acs{RT} dispatch tracking, demonstrating that the adaptive lower control layer primarily delivers benefits through deviation mitigation.

\begin{figure}[pos=ht]
    \centering
    \begin{subfigure}[t]{\linewidth}
    \centering
    \includegraphics[width=\linewidth]{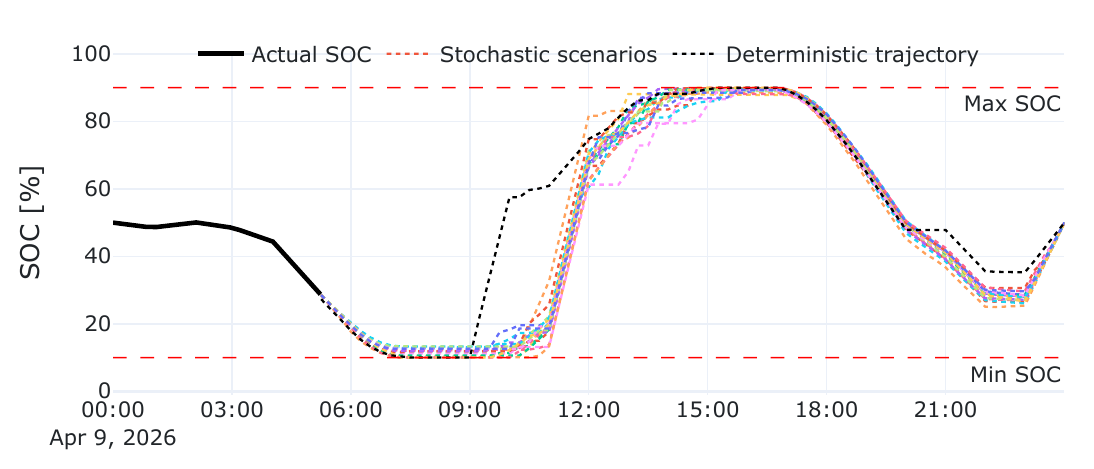}
    \caption{Trajectories at 05:15.}   
    \end{subfigure}
    ~
    \begin{subfigure}[t]{\linewidth}
    \centering
    \includegraphics[width=\linewidth]{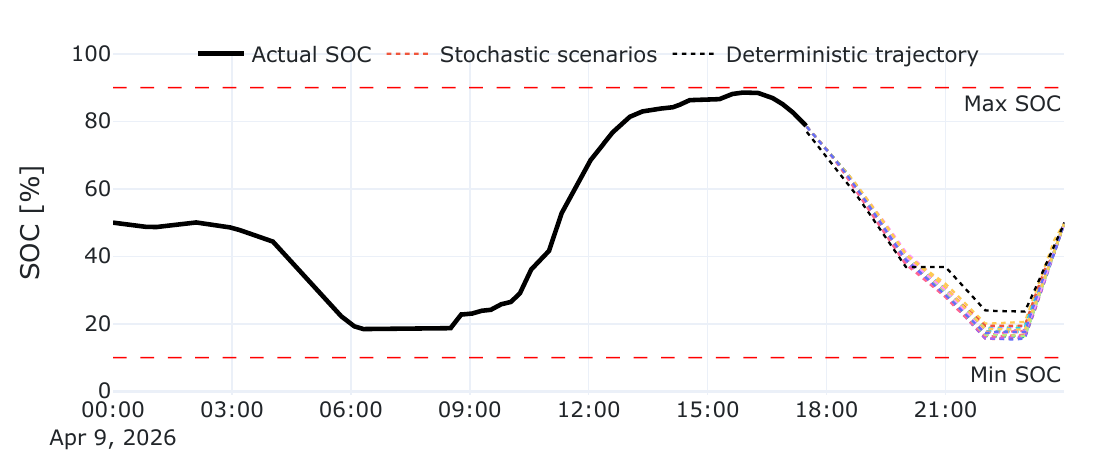}
    \caption{Trajectories at 17:30.}   
    \end{subfigure}
    \caption{Planned \acs{BESS} SOC trajectories of Apr.9, 2026.}
    \label{fig-bess-prediction-2026-04-09}
\end{figure}

The planned \acs{BESS} SOC trajectories predicted at 05:15 and 17:30 are shown in \Cref{fig-bess-prediction-2026-04-09}. The bold black curve denotes the realized SOC, the black dashed curve corresponds to the deterministic controller, and the colored dashed curves represent the stochastic scenarios considered by the upper-layer \acs{MPC}. All trajectories start from an initial SOC of \qty{50}{\percent} and are scheduled to return to the same level at the end of the day while remaining within the operational limits of \qty{10}{\percent} and \qty{90}{\percent}. Compared with the deterministic controller, the stochastic \acs{MPC} explicitly accounts for multiple uncertainty realizations during planning. Consequently, the charging strategy avoids relying on optimistic forecasts and maintains feasible operating trajectories across a wide range of future scenarios, providing improved robustness against \acs{PV} generation and workload uncertainties.

\subsection{Day II: Overcast winter conditions}

\begin{figure}[pos=ht]
    \centering
    \includegraphics[width=\linewidth]{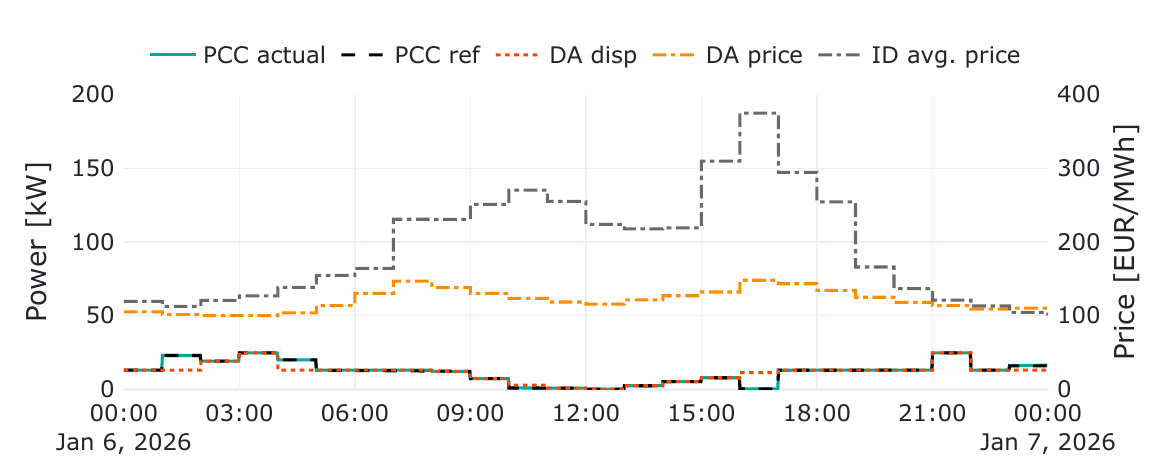}
    \caption{\acs{DA} plan, \acs{ID}-adjusted dispatch reference, realized \acs{PCC} power, and market prices of Jan.6, 2026.}
    \label{fig-pcc-da-disp-2026-01-06}
\end{figure}

\begin{figure}[pos=ht]
    \centering
    \includegraphics[width=\linewidth]{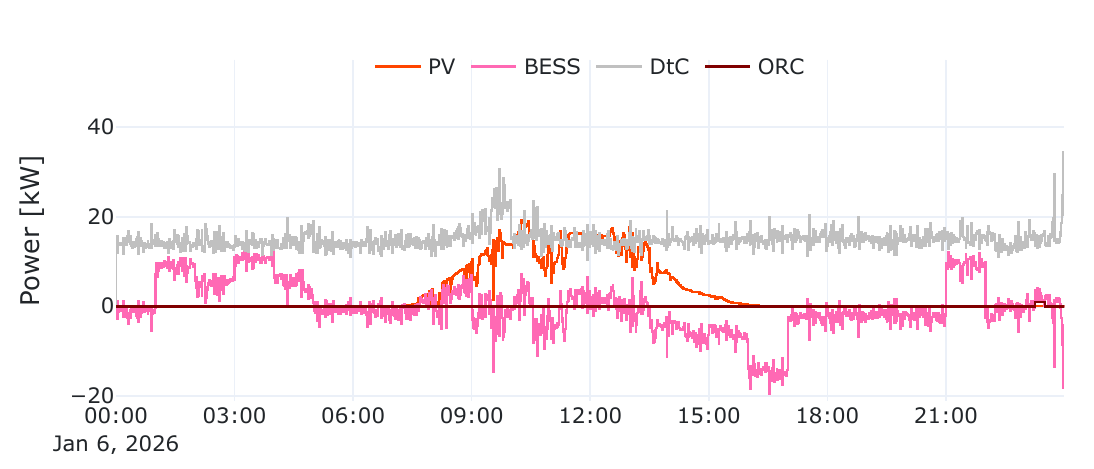}
    \caption{System operation over the day of Jan.6, 2026.}
    \label{fig-operation_2026-01-06_two_layer_stochastic}
\end{figure}

January 6th, 2026 is selected as a representative overcast winter day to evaluate the proposed framework under conditions of limited on-site generation. In contrast to the clear-sky case, persistent cloud cover considerably reduces the available \acs{PV} generation, resulting in a much stronger reliance on electricity imported from the public grid, as shown in \Cref{fig-pcc-da-disp-2026-01-06} and \Cref{fig-operation_2026-01-06_two_layer_stochastic}. Consequently, the opportunity for shifting batch workloads toward periods of abundant \acs{PV} generation becomes limited, and the operation is primarily driven by electricity market conditions. Since the system operates during the heating season, recovered waste heat is supplied to the district heating network instead of being converted into electricity, making the \acs{ORC} unit inactive throughout the day. Under these operating conditions, the \acs{BESS} remains the primary source of operational flexibility, while its role shifts from maximizing renewable self-consumption to exploiting electricity price variations through energy arbitrage.


\Cref{fig-pcc-da-disp-2026-01-06} also illustrates the interaction with the electricity market. Similar to Day~I, the realized \acs{PCC} power closely follows the \acs{ID}-adjusted dispatch reference throughout the day. However, unlike the renewable-driven operation observed under clear-sky conditions, the charging and discharging schedule of the \acs{BESS} closely follows electricity price variations, demonstrating that market arbitrage becomes the dominant operational objective when local renewable resources are limited.

\paragraph{Case study: high carbon price}

To investigate the influence of carbon-aware operation, the baseline carbon price of $\qty{265}{EUR/tCO_2eq}$, corresponding to the estimated social cost of carbon reported in \cite{rennertComprehensiveEvidenceImplies2022}, is increased to $\qty{1000}{EUR/tCO_2eq}$, representing a conservative estimate of the current cost of direct air capture technologies \cite{sievertConsideringTechnologyCharacteristics2024}.

\begin{figure}[pos=htbp]
    \centering
    \includegraphics[width=\linewidth]{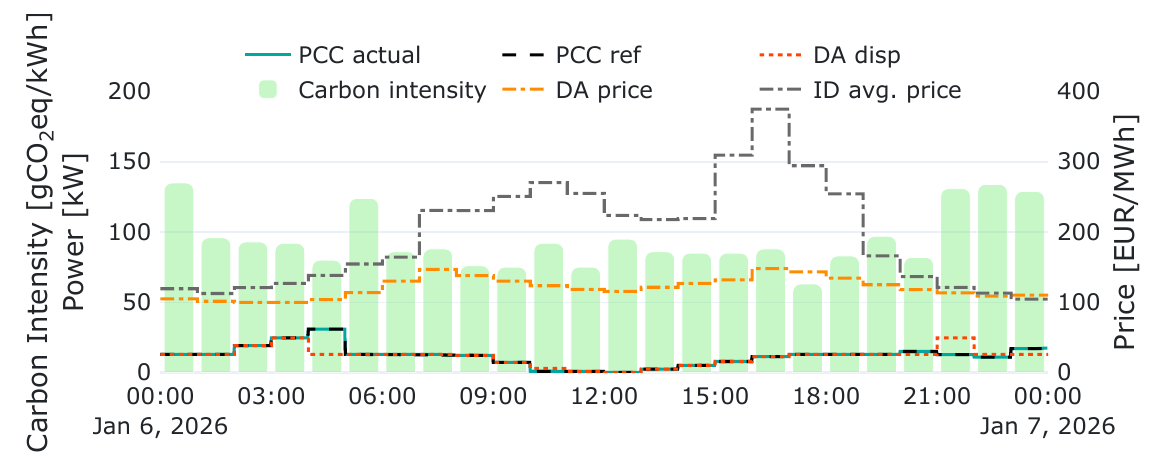}
    \caption{\acs{DA} plan, \acs{ID}-adjusted dispatch reference, realized PCC power, market prices, and grid carbon intensity of Jan.6, 2026 under high carbon price.}
    \label{fig-pcc_da_disp_2026-01-06_two_layer_stochastic_high_carbon}
\end{figure}

The resulting operation schedule is shown in \Cref{fig-pcc_da_disp_2026-01-06_two_layer_stochastic_high_carbon}. Under the increased carbon price, the controller actively avoids importing electricity during periods with high grid carbon intensity (e.g. 21:00). This change
illustrates that the proposed framework is capable of explicitly incorporating
carbon pricing into market-oriented scheduling decisions while simultaneously
accounting for economic objectives. For the selected winter day, this behavioral change results in an
approximately $\qty{1}{kgCO_2eq}$ ($\approx3\%$) reduction in emissions
associated with grid electricity imports. Although the absolute reduction is
marginal for the considered operating conditions, it reflects the limited flexibility available on an overcast winter day with little local
\acs{PV} generation. Larger emission reductions are expected under operating
conditions offering greater scheduling flexibility or stronger temporal
variations in external grid carbon intensity.

\begin{figure}[pos=ht]
    \centering
    \includegraphics[width=\linewidth]{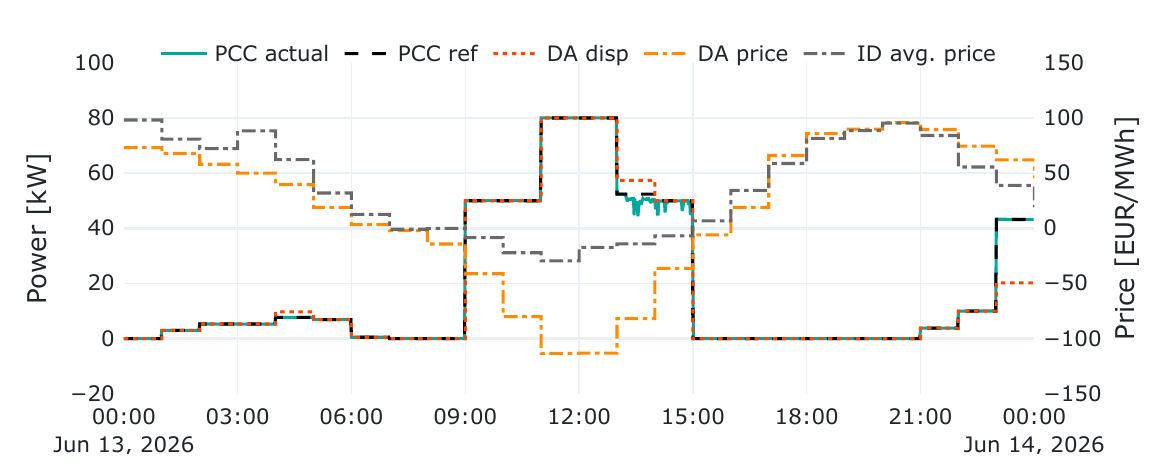}
    \caption{\acs{DA} plan, \acs{ID}-adjusted dispatch reference, realized \acs{PCC} power, and market prices of Jun.13, 2026.}
    \label{fig-pcc-da-disp-2026-06-13}
\end{figure}

\begin{figure}[pos=ht]
    \centering
    \includegraphics[width=\linewidth]{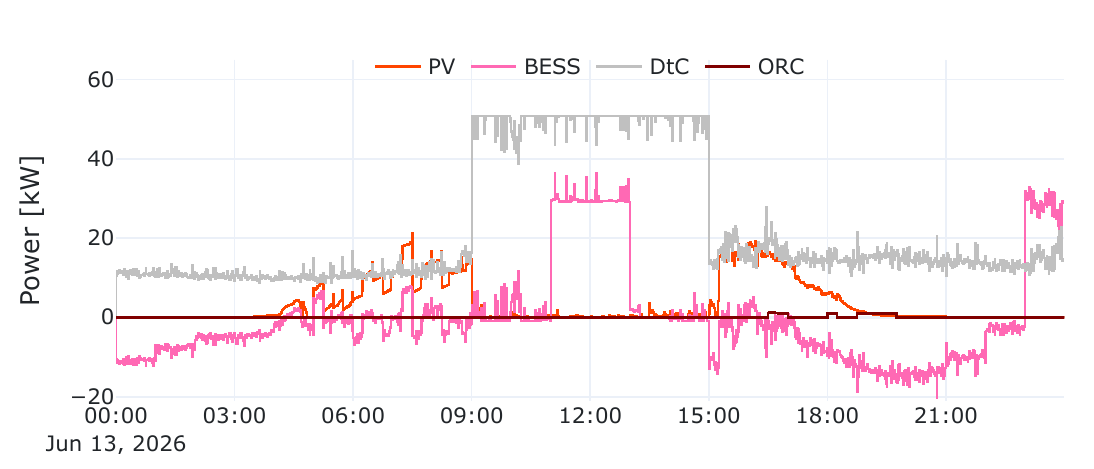}
    \caption{System operation over the day of Jun.13, 2026.}
    \label{fig-operation_2026-06-13}
\end{figure}

{
\subsection{Day III: Summer oversupply conditions}\label{sec-day-III}

To further evaluate the adaptability of the proposed framework under extreme seasonal and market conditions, June 13th, 2026 is selected as a summer day characterized by negative electricity prices. As shown in \Cref{fig-pcc-da-disp-2026-06-13}, the \acs{DA} dispatch plan responds to the price signal by increasing electricity consumption during negative-price periods. Together with \Cref{fig-operation_2026-06-13}, the results show that the proposed framework coordinates multiple flexibility resources accordingly: the \acs{BESS} is charged, batch workloads are shifted toward periods with negative prices, \acs{PV} generation is curtailed when additional generation is economically unfavorable, and the \acs{ORC} remains inactive. The realized \acs{PCC} power closely follows the \acs{ID}-adjusted reference, demonstrating the ability of the framework to adapt both its scheduled operation and resource dispatch to favorable market conditions.
}


\paragraph{Case study: external benchmark}

\begin{figure}[pos=ht]
    \centering
    \includegraphics[width=\linewidth]{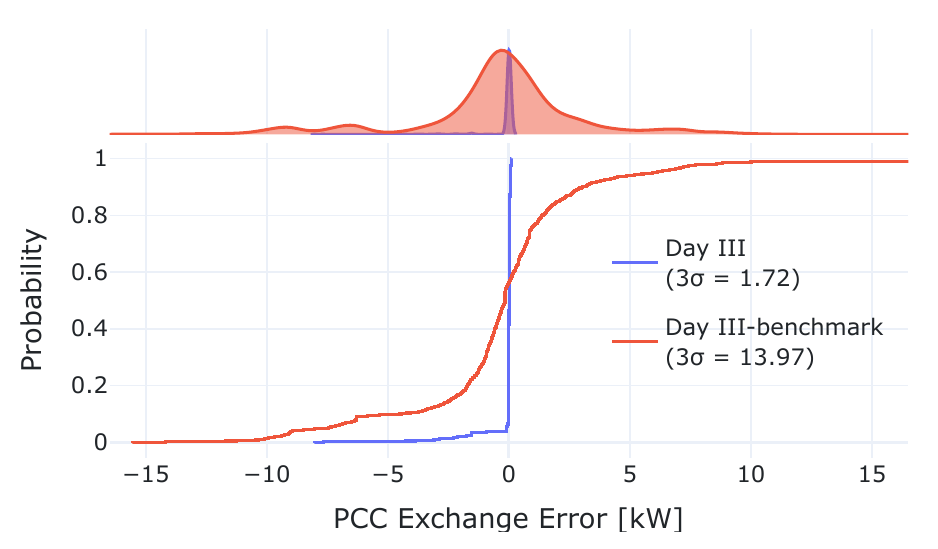}
    \caption{Empirical distribution of power tracking errors at \acs{PCC} for Day III and the external benchmark.}
    \label{fig-pcc-error-ecdf-2026-06-13}
\end{figure}

\begin{figure}[pos=ht]
    \centering
    \includegraphics[width=\linewidth]{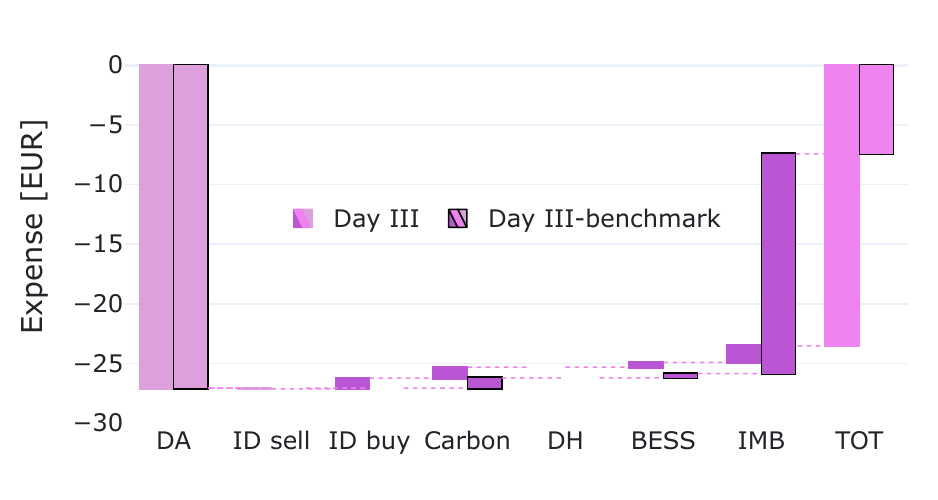}
    \caption{Ex post operational expense breakdown of Day III and the benchmark.}
    \label{fig-expense-breakdown-benchmark}
\end{figure}

To provide an external benchmark, the proposed framework is compared with an adapted hierarchical \acs{MPC} strategy based on
\cite{guoMultitimescaleOptimizationScheduling2024}. The benchmark retains the same \acs{DtC} power model and physical system, but uses a deterministic \qty{1}{h} upper-layer scheduling resolution and a \qty{15}{min} lower-layer control resolution, without active \acs{ID} bidding or tube-based \acs{RT} disturbance compensation. \Cref{fig-pcc-error-ecdf-2026-06-13} compares the empirical distribution of \acs{PCC} tracking errors, while \Cref{tab-day-iii-metrics} summarizes the corresponding tracking and economic performance. The proposed framework achieves tighter tracking, reducing the \acs{MAE} from \qty{2.502}{kW} to \qty{0.119}{kW} and the {RMSE} from \qty{4.657}{kW} to \qty{0.578}{kW}, with the $3\sigma$ error range decreasing from \qty{13.965}{kW} to \qty{1.715}{kW}. As shown in \Cref{fig-expense-breakdown-benchmark}, both strategies benefit from the negative electricity prices observed on Day~III, resulting in net operational revenues. However, the larger tracking deviations of the benchmark are accompanied by much higher ex-post imbalance expenses, reducing its net operational revenue. Overall, the proposed framework achieves a lower total expense of compared with the benchmark case. It should be noted that the two approaches differ in several aspects, including uncertainty treatment, scheduling resolution, \acs{ID} participation, and \acs{RT} compensation. Therefore, this external comparison evaluates the combined benefits of the proposed framework rather than isolating individual contributions, which are examined through the internal comparisons in \Cref{sec-day-I}.

\begin{table}[pos=ht]
\centering
\caption{Summary of key metrics for Day III and Day III-benchmark.}
\label{tab-day-iii-metrics}
\begin{tabular}{lcc}
\toprule
Metric & Day III & Day III-benchmark \\
\midrule
DA expense [EUR] & -27.065 & -27.065 \\
ID trading expense [EUR] & 0.841 & 0.000 \\
Imbalance expense [EUR] & 1.388 & 18.393 \\
Other expenses [EUR] & 1.317 & 1.229 \\
Tot. expense [EUR] & -23.519 & -7.444 \\
PCC tracking MAE [kW] & 0.119 & 2.502 \\
PCC tracking RMSE [kW] & 0.578 & 4.657 \\
PCC tracking $3\sigma$ [kW] & 1.715 & 13.965 \\
Tot. Imb. energy [kWh] & 2.776 & 36.786 \\
\bottomrule
\end{tabular}
\end{table}

\subsection{Discussion and comparative analysis}

\begin{figure}
    \centering
    \includegraphics[width=.85\linewidth]{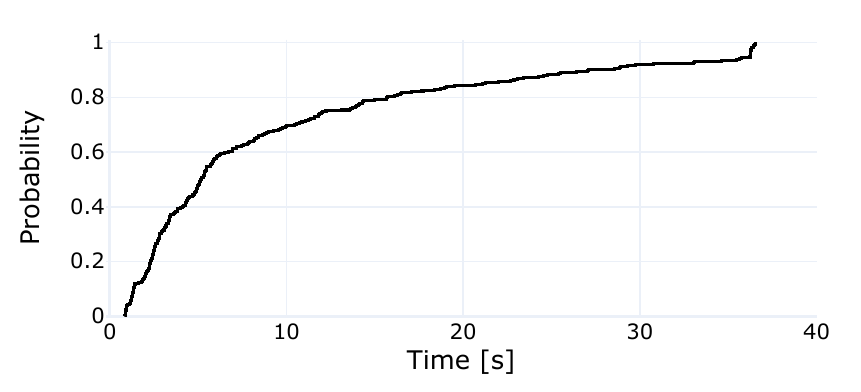}
    \caption{ECDF plot of the upper layer computation time.}
    \label{fig-upper-timer-ecdf}
\end{figure}

The proposed control framework is implemented on a workstation equipped with an Apple M5 Pro processor and \qty{48}{GB} of RAM. {The optimization problems are formulated using Linopy v0.8.0 \cite{hofmannLinopyLinearOptimization2023} and solved using the commercial solver Gurobi v13.0 \cite{gurobi}. A solver time limit of \qty{35}{s} is imposed on the upper-layer optimization to bound the computational effort during \acs{RT} operation.} The computation times of the upper- and lower-layer controllers for the baseline cases of Day~I to Day~III are summarized in \Cref{tab-computation-time}. {The lower-layer optimization is consistently solved within \qty{2}{s}, whereas the complete upper-layer control loop, including forecasting, scenario generation and reduction, optimization model construction, and solution, requires at most \qty{36.6}{s}. The slightly larger end-to-end computation time results from the additional preprocessing and model construction overhead. {The empirical
cumulative distribution of upper-layer computation times is shown in
\Cref{fig-upper-timer-ecdf}, indicating that the majority of control
cycles are completed well below the worst-case value.} Since the upper- and lower-layer controllers operate at sampling intervals of \qty{15}{min} and \qty{1}{min}, respectively, the measured computation times provide large margins for their respective \acs{RT} requirements.}

\begin{table}[pos=ht]
    \centering
    \caption{Computation time statistics for the two control layers.}\label{tab-computation-time}
    \begin{tabular}{cccc}
    \toprule
        Control layer & Min [s] & Average [s] & Max [s] \\
    \midrule
         Upper& 0.85 & 9.61 & 36.6 \\ 
         Lower& 0.28 & 0.31 & 1.67 \\
    \bottomrule
    \end{tabular}
\end{table}

\begin{figure}[pos=ht]
    \centering
    \includegraphics[width=\linewidth]{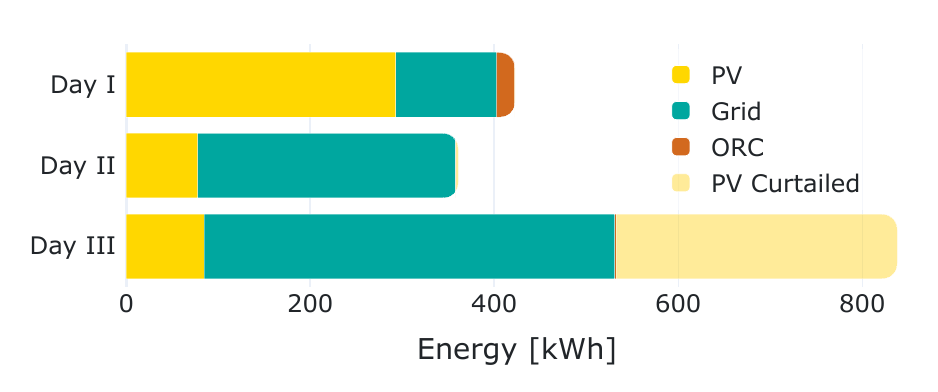}
    \caption{Energy mix comparison between the two representative operating days.}
    \label{fig-mix-comparison}
\end{figure}

\Cref{fig-mix-comparison} compares the energy supply composition under the three representative operating conditions. During the clear-sky day, on-site \acs{PV} generation supplies approximately \qty{300}{kWh} of electricity, with an additional contribution of about \qty{20}{kWh} recovered through the \acs{ORC} turbine. In contrast, the overcast winter day exhibits substantially lower renewable production, requiring higher electricity imports from the public grid. {Under the negative-price conditions of Day~III, the operating strategy changes accordingly: electricity consumption is largely shifted toward negatic-price periods, while part of the \acs{PV} generation is curtailed when additional electricity production is economically unfavorable.} These results illustrate the ability of the proposed framework to adapt its resource dispatch to both seasonal renewable availability and market conditions.

\begin{figure}[pos=ht]
    \centering
    \includegraphics[width=\linewidth]{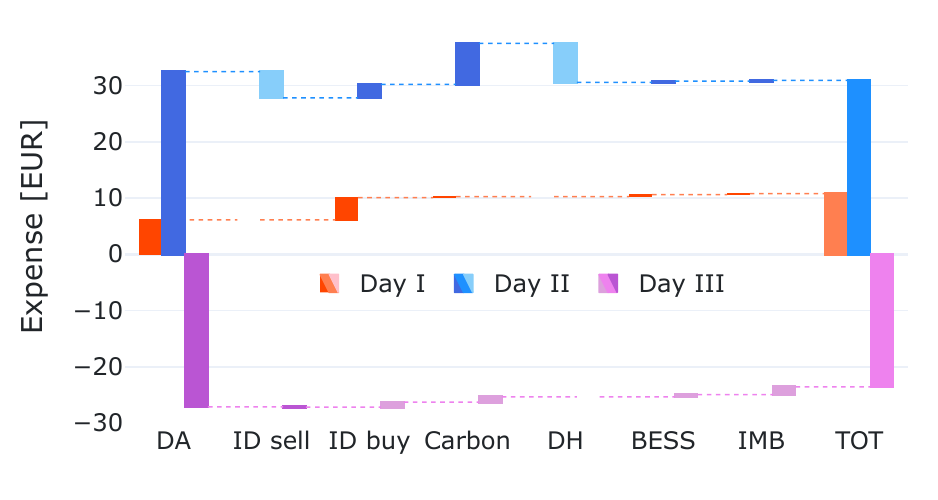}
    \caption{Ex post operational expense breakdown under different operating conditions.}
    \label{fig-expense-breakdown}
\end{figure}

The overall ex post operational expenses are summarized in \Cref{fig-expense-breakdown}. The overcast winter case results in substantially higher operating costs, primarily due to increased \acs{DA} electricity procurement. {In contrast, the negative-price conditions of Day~III allow the available flexibility to be exploited for economic benefit, resulting in a negative net operational expense (revenue). Across all cases, battery degradation costs remain relatively marginal.} Although the case study is based on a laboratory-scale system, the proposed optimization framework is generic and can be readily scaled to larger installations.

\section{Conclusion}

This paper proposed a two-layer \acs{RT} control framework for sustainable \acs{DtC} operation in on-site multi-energy systems. The framework combines a stochastic upper-layer \acs{MPC} for \acs{ID} market participation and energy scheduling with an adaptive tube-based lower-layer \acs{MPC} that compensates short-term uncertainties and promotes \acs{RT} dispatch tracking. The control architecture explicitly coordinates \acs{DtC} workload flexibility, \acs{BESS}, \acs{PV} generation, waste heat recovery, and \acs{DH} while incorporating forecasts of on-site \acs{RES}, electricity prices, workload arrivals, and grid carbon intensity.

Simulation studies under representative operating conditions demonstrated the effectiveness of the proposed framework. The results showed that the adaptive lower control layer substantially improved tracking of the \acs{ID} dispatch reference compared with single-layer control strategies, thereby reducing imbalance costs caused by forecast and model errors. The stochastic upper layer successfully coordinated flexible workloads and energy resources under uncertainty, while naturally adapting the operating strategy to seasonal changes in \acs{RES} and electricity prices. Furthermore, the carbon-price sensitivity study showed that the proposed framework can translate carbon-aware economic incentives into operational decisions, shifting energy usage toward periods with lower grid carbon intensity and reducing emissions associated with imported electricity. The proposed framework is computationally tractable for \acs{RT} application and provides a practical pathway toward economically efficient, carbon-aware, and grid-supportive operation of future sustainable \acs{DtC}s.

Beyond the presented case studies, the proposed framework is intended for \acs{DtC} operators and energy management system providers seeking to integrate flexible computing workloads with \acs{DER}s and electricity market participation. Its modular architecture enables the incorporation into existing energy management systems and makes it a practical and scalable solution for future sustainable \acs{DtC}s operating under increasingly stringent grid integration requirements. Future work will focus on extending the market participation strategy to jointly optimize \acs{DA}, \acs{ID}, and balancing markets, as well as validating the framework on a physical testbed. Additional research directions include improving forecasting accuracy through advanced learning-based models and incorporating more detailed carbon intensity estimation methods that capture the spatial and temporal variability of electricity emissions at the distribution network level.









\appendix

{
\section{Appendix: comparison of queuing models} \label{sec-app-queuing-model-comparison}

To assess the approximation introduced by the independent M/M/1 model, its response time is compared with that of an M/M/c queue over the workload range considered in this study. For a queue with $c$ parallel servers, aggregate arrival rate $\lambda$, and per-server service rate $\mu$, the M/M/c model with an ideal load balancer gives the mean response time \cite{gautam2012analysis} in \eqref{eq-app-mmc}:
\begin{equation}
t_{\mathrm{M/M/c}}
=
\frac{\mathrm{C}(c,\lambda,\mu)}
{c\mu-\lambda}
+
\frac{1}{\mu},
\label{eq-app-mmc}
\end{equation}
where $\mathrm{C}(c,\lambda,\mu)$ denotes the Erlang C probability of an arriving request having to wait in the queue.

For comparison, two M/M/1 representations provide useful reference points. Treating the $c$ servers as a single equivalent server with service rate $c\mu$ gives
\begin{equation}
t_{\mathrm{M/M/1,scaled}}
=
\frac{1}{c\mu-\lambda},
\label{eq-app-sc-mm1}
\end{equation}
whereas assigning the aggregate arrival rate uniformly across independent servers gives
\begin{equation}
t_{\mathrm{M/M/1,independent}}
=
\frac{1}{\mu-\lambda/c}.
\label{eq-app-id-mm1}
\end{equation}

\Cref{fig-queue-model} compares the three models using the server and workload parameters of the testing framework. The independent M/M/1 approximation consistently yields a higher response-time estimate than the M/M/c model over the considered operating region, while the scaled M/M/1 model provides a lower estimate. Thus, the selected approximation introduces a conservative \acs{QoS} margin relative to the M/M/c representation.
\begin{figure}[pos=htbp]
    \centering
    \includegraphics[width=\linewidth]{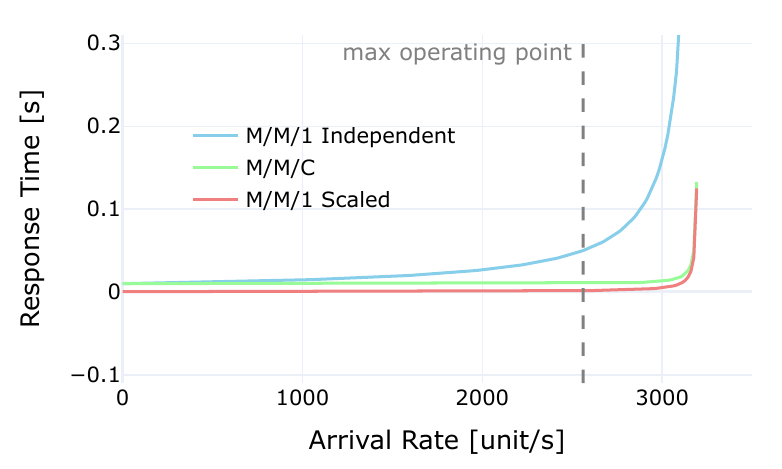}
    \caption{The queuing model comparison.}
    \label{fig-queue-model}
\end{figure}
Since the queuing model is used only to exclude operating points that violate the latency requirement during the construction of the \acs{DtC} power model, this conservative approximation is considered sufficient for the purpose of the proposed control framework.

}

\section{Appendix: stochastic scenario reduction} \label{sec-app-stochastic-scenarios}

{\subsection{Scenario generation}}
Uncertainties in the upper-layer \acs{MPC} are modeled using scenario-based stochastic programming. The stochastic variables considered include \acs{GHI} {where the \acs{DtC} is located}, interactive { computing} workload, and \acs{ID} electricity prices, with their forecasts obtained from the forecasting framework described in Appendix \ref{sec-app-forecasting}. The carbon intensity forecast is treated deterministically since the provider currently supplied only point predictions \cite{electricitymaps2025data}.

{The stochastic interactive workload represents external computational demand such as world wide web services (i.e. from geographically distributed user requests), rather than local weather-dependent thermal or cooling loads. It is therefore treated as independent of both local meteorological conditions and electricity market signals. To assess potential dependence between local \acs{GHI} and \acs{ID} prices, we remove their common temporal structure by considering the \acs{ID} price residual relative to the corresponding \acs{DA} price. The Pearson correlation coefficient is then evaluated using the 2024 data. As shown in \Cref{tbl-ghi-id-corr}, the residual correlation between local \acs{GHI} and \acs{ID} price fluctuations is weak, indicating limited dependence between these two uncertainty sources after accounting for their predictable temporal structure.}

\begin{table}[pos=ht]
    \centering
    \caption{The Pearson correlation coefficient between \acs{GHI} and \acs{DA}-\acs{ID} price residues}
    \label{tbl-ghi-id-corr}
    \begin{tabular}{ccccc}
    \toprule
    & \acs{GHI} & \acs{ID} low & \acs{ID} mean & \acs{ID} high \\
    \midrule
    \acs{GHI} & 1.000 & -0.114 & -0.067 & 0.018 \\
    \acs{ID} low & -0.114 & 1.000 & 0.634 & 0.096 \\
    \acs{ID} mean & -0.067 & 0.634 & 1.000 & 0.431 \\
    \acs{ID} high & 0.018 & 0.096 & 0.431 & 1.000 \\
    \bottomrule
    \end{tabular}
\end{table}

{Based on these observations, the forecast errors of the considered stochastic variables are assumed to be independent\footnote{The scenario-based formulation can be extended to jointly sampled scenarios if statistically significant cross-variable dependence is identified.}}. Raw scenarios are therefore generated separately for each variable at every \acs{MPC} iteration using the latest available forecasts, producing full-horizon trajectories with uncertainty increasing with the prediction horizon.

{\subsection{Scenario reduction}}
To avoid the combinatorial growth caused by enumerating all joint realizations, a Monte Carlo sampling approach is adopted. A fixed number of combined scenarios is constructed by randomly sampling and concatenating raw trajectories of irradiance, workload, and electricity prices into multivariate scenario paths. Buy and sell price trajectories are sampled jointly to preserve consistency. To maintain computational tractability within the \acs{MPC} loop, a forward-selection scenario reduction algorithm proposed in \cite{dupacovaScenarioReductionStochastic2003, heitschScenarioTreeModeling2009} is employed. Each combined scenario is represented as a normalized multivariate trajectory over the prediction horizon. The distance between two scenarios is defined as:
\begin{equation}
    d(i,j) = \sum_{k=\bar{k}}^{T} w_k \, \lVert \tilde{s}_i(k) - \tilde{s}_j(k) \rVert_2
    \label{eq-distance-scenario-ij}
\end{equation}
where $\tilde{s}_i(k)$ denotes the normalized scenario vector and $w_k=\alpha^{k-1}$ is a temporal weighting factor that emphasizes near-term uncertainty. Starting from an empty reduced set, scenarios are iteratively selected to minimize the aggregate distance (\Cref{alg-forward-selection}), and the probability of each representative scenario is obtained from the proportion of original scenarios assigned to it (\Cref{alg-probability-assignment}). The final reduced scenario set contains $N_{\text{reduced}}$ representative trajectories.

\begin{algorithm}[ht]
\caption{Greedy forward selection}
\label{alg-forward-selection}
\begin{algorithmic}[1]
\REQUIRE Set of combined scenarios $\{\xi^0,\dots,\xi^{N_\text{comb}}\}$, target number of scenarios $N_\text{reduced}$, distance metric $d(\cdot,\cdot)$
\ENSURE Reduced scenario index set $\mathcal{J}$

\STATE Compute pairwise distance matrix, use the distance metric defined in \Cref{eq-distance-scenario-ij}:
\[
D_{i\!j} = d(\xi^i,\xi^j), \quad \forall i,j = 0,\dots,N_\text{comb}-1
\]

\STATE Initialize reduced set $\mathcal{J} \leftarrow \emptyset$
\STATE Initialize distances to reduced set:
\[
\delta_i \leftarrow +\infty, \quad \forall i = 0,\dots,N_\text{comb}-1
\]

\FOR{$n = 0$ to $N_\text{reduced}-1$}
    \FOR{each candidate $k \notin \mathcal{J}$}
        \STATE Compute marginal gain:
        \[
        \Delta(k) = \sum_{i=0}^{N_\text{comb}-1} \left( \delta_i - \min(\delta_i, D_{ik}) \right)
        \]
    \ENDFOR
    \STATE Select scenario: $k^* = \arg\max_{k \notin \mathcal{J}} \Delta(k)$
    \STATE Update reduced set: $ \mathcal{J} \leftarrow \mathcal{J} \cup \{k^*\}$
    \STATE Update distances:
    \[
    \delta_i \leftarrow \min(\delta_i, D_{ik^*}), \quad \forall i = 0,\dots,N_\text{comb}-1
    \]
\ENDFOR

\RETURN $\mathcal{J}$
\end{algorithmic}
\end{algorithm}

\begin{algorithm}[ht]
\caption{Probability assignment for reduced scenarios}
\label{alg-probability-assignment}
\begin{algorithmic}[1]
\REQUIRE Distance matrix $D_{ij}$, reduced scenario indices $\mathcal{J}$
\ENSURE Scenario probabilities $\{p_j\}_{j \in \mathcal{J}}$

\FOR{each original scenario $i = 1,\dots,M$}
    \STATE Assign nearest representative: $j^*(i) = \arg\min_{j \in \mathcal{J}} D_{i\!j}$
\ENDFOR

\FOR{each reduced scenario $j \in \mathcal{J}$}
    \STATE Compute probability: $p_j = {|\{ i \mid j^*(i) = j \}|}/{M}$
\ENDFOR

\RETURN $\{p_j\}_{j \in \mathcal{J}}$
\end{algorithmic}
\end{algorithm}

{\subsubsection{Sensitivity to the number of reduced scenarios}}
\label{sec-app-scenario-number-sensitivity}

\begin{figure}[pos=htbp]
    \centering
    \includegraphics[width=\linewidth]{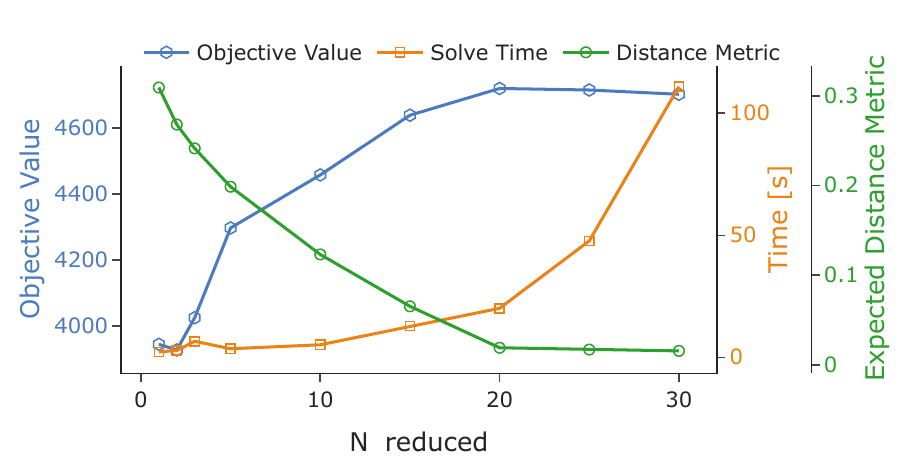}
    \caption{Sensitivity of the upper-layer optimization to the number of reduced scenarios.}
    \label{fig-scenario-number-sensitivity}
\end{figure}

{
The number of reduced scenarios, $N_{\mathrm{reduced}}$, determines the trade-off between the fidelity of the stochastic representation and the computational burden of the upper-layer \acs{MPC}. To quantify this trade-off, a sensitivity analysis is performed at the first control step of Day I presented in \Cref{sec-day-I}, using the same forecasts and raw scenario set as in the corresponding simulation. The optimal objective value, the scenario-reduction distance metric \eqref{eq-distance-scenario-ij}, and the computation time are evaluated for different values of $N_{\mathrm{reduced}}$. The solver time limit used in the \acs{RT} implementation is relaxed in this analysis to characterize the convergence behavior as $N_{\mathrm{reduced}}$ increases. As shown in \Cref{fig-scenario-number-sensitivity}, increasing $N_{\mathrm{reduced}}$ progressively decreases the distance metric, indicating a closer approximation of the original scenario distribution. The optimal objective value also approaches a stable value as the scenario-set size increases, while the computation time rises substantially. In particular, the objective value and distance metric exhibit only limited changes beyond approximately $N_{\mathrm{reduced}}=20$, whereas further increasing the number of scenarios results in a considerably larger computational burden. Accordingly, $N_{\mathrm{reduced}}=20$ is adopted in the subsequent simulations as a practical compromise between stochastic representation accuracy and computational efficiency.
}

\section{Appendix: results of the linearization of the data center model}\label{sec-app-linearization}

\begin{table*}[ht]
\centering
\caption{Details of the region split for linearization}
\label{tab-linearization}
\renewcommand{\arraystretch}{1.5}
\begin{tabular}{|c|c|c|c|}
\hline
\textbf{i} & \textbf{Conditions} & $w^i_\text{lb}$ [W] &  $w^i_\text{ub}$ [W] \\ \hline
\textbf{1} & 
$t_D \cdot b \le  N \  \land \  b\cdot (0.5 \mu t_D  - 1) \le a \  \land \ b + 4a \le 1.7  N \mu \  \land \  a + b \ge N/t_D$ & -925.4 & 364.7 \\ \hline

\textbf{2} & 
$t_D \cdot b \le  N \  \land \  t_D \cdot a \le  N(\mu t_D  - 1)\  \land \ b + 4a \ge 1.7  N \mu \  \land \  a + b \le 0.6  N \mu$ & -531.8 & 248.7 \\ \hline

\textbf{3} & 
$t_D \cdot b \le  N \  \land \  t_D \cdot a \le  N(\mu t_D  - 1)\  \land \ b + 4a \ge 1.7  N \mu \  \land \  a + b \ge 0.6 \cdot  N \mu\ \land \ a+b \le 0.875 N \mu $ & -2671.0 & 984.8 \\ \hline

\textbf{4} & 
$t_D \cdot b \ge  N \  \land \  a + b \le  N \mu \  \land \ a + b \ge  0.5 N \mu \  \land \  a \le 0.6  N \mu\ \land \ a+b \le 0.875 N \mu $ & -2137.7 & 1123.2 \\ \hline

\textbf{5} & 
$t_D \cdot b \ge  N \  \land \  a + b \le  N \mu \  \land \ a + b \ge  0.5 N \mu \  \land \  a \ge 0.6   N \mu\ \land \ a+b \le 0.875 N \mu $  & -88.4 & 49.8 \\ \hline

\textbf{6} & 
$b\cdot(0.5 \mu t_D - 1) \ge a \  \land \  a + b \le  0.5 N \mu \  \land \ a + b \ge N/t_D\ \land \ a+b \le 0.875 N \mu $  & -892.0 & 497.8 \\ \hline

\textbf{7} & 
$a + b \le N/t_D$  & -197.6 & 114.5 \\ \hline
\textbf{8} & 
$a+b \ge 0.875 N \mu$  & -1651.5 & 267.5 \\ \hline
\end{tabular}
\end{table*}

The workload domain is partitioned into $N_{\text{region}} = 8$ regions based on the conditions shown in \Cref{tab-linearization}. The resulting linear coefficients, intercepts and their respective $R^2$ values are detailed in \eqref{eq-p-dtc-fit-results}. The
linearization error bounds obtained for each region are also summarized in
\Cref{tab-linearization}.
\begin{equation}
\begin{cases}
\hat{\varphi}_1(a,b) = 4488.27 + 7.22a + 2.98b      & R^2 = 0.9728\\
\hat{\varphi}_2(a,b) = -6647.26 + 15.53a + 5.62b    & R^2 = 0.9946 \\
\hat{\varphi}_3(a,b) = -23918.58 + 25.01a + 9.39b  & R^2 = 0.9842 \\
\hat{\varphi}_4(a,b) = -19194.12 + 18.82a + 18.8b  & R^2 = 0.9777 \\
\hat{\varphi}_5(a,b) = -43222.73 + 28.23a + 28.2b  & R^2 = 0.9994 \\
\hat{\varphi}_6(a,b) = 4953.35 + 4.98a + 4.65b      & R^2 = 0.9391 \\
\hat{\varphi}_7(a,b) = 8019.25 + 1.85a + 0.70b      & R^2 = 0.9540 \\
\hat{\varphi}_8(a,b) = -69125.83 + 37.41a + 37.39b      & R^2 = 0.9975 \\
\end{cases} \label{eq-p-dtc-fit-results}
\end{equation}
{To assess the benefit of the regional decomposition, a single-region affine
model \eqref{eq-p-dtc-fit-results-single} is considered as a baseline:
\begin{equation}
    \hat{\varphi}_{\rm single}(a,b)
    = 8019.25 + 9.39a + 8.68b,
    \quad R^2 = 0.6961
    \label{eq-p-dtc-fit-results-single}
\end{equation}
This baseline represents a conventional global linear approximation in which
the nonlinear \acs{DtC} power characteristics are described by a single
affine relationship over the entire workload domain. 

The validation performance of the two models is summarized in \Cref{tab-dtc-model-validation}. The proposed piecewise-linear model achieves substantially lower errors, by one order of magnitude, than the single-region baseline, with an \acs{MAE} of \qty{425}{W} compared with \qty{6050}{W} for the baseline. Together with the reported $R^2$ values, these results demonstrate that the piecewise representation provides a sufficiently accurate approximation and narrows the uncertainty bounds used by the adaptive tube controller.
}

\begin{table}[pos=ht]
    \centering
    \caption{Validation performance of the linearized \acs{DtC} power models.}
    \label{tab-dtc-model-validation}
    \begin{tabular}{cccc}
    \toprule
        Model & MAE [W] & RMSE [W] & MAPE [\%] \\
    \midrule
        Single-region & 6050 & 6871 & 24.9 \\
        Piecewise linear & 425 & 626 & 2.28 \\
    \bottomrule
    \end{tabular}
\end{table}

\section{Appendix: forecasting} \label{sec-app-forecasting}

The two control layers operate at different time scales and require different forecast inputs. Interactive workload and \acs{GHI} forecasts are used by both layers, whereas \acs{ID} price and carbon intensity forecasts are only required by the upper-layer optimization for \acs{ID} market participation and carbon-aware scheduling.


\subsection{Interactive workload forecasting} \label{sec-app-forecasting-short-workload}

A linear regression model is trained to the medium-term (over a horizon of up to 24 hours with a temporal resolution of 15 minutes) interactive workload arrivals with the lagged workload features as explanatory variables: workload lagged by one week, two days, and one day. The regression model produces both point forecasts and corresponding 90\% \acs{CI}s.

Since the minute-scale workload fluctuations exhibit little temporal correlation after removing the medium-term forecast, short-term forecasting is not performed explicitly. Instead, the residual uncertainty is estimated online and used by the lower-layer tube-based \acs{MPC}. At each control step, the prediction error between the measured workload and the medium-term forecast is used to update an exponentially weighted estimate of the residual variance, which assigns higher weight to recent observations, enabling the controller to track changes in workload volatility in real time. The resulting variance estimation is used to construct the uncertainty bounds required by the lower control layer.

\subsection{GHI forecasting}
\subsubsection{Short-term GHI forecasting} \label{sec-app-forecasting-short-ghi}
A short term forecasting method from \cite{scolariIrradiancePredictionIntervals2016, scolariModellingForecastingPhotovoltaic2019a} is used in this work to forecast the \acs{GHI} intervals, which uses a nonparametric approach to construct \acs{PI}s by leveraging a pattern recognition technique. The fundamental principle is that, if a specific pattern (average clear-sky index and its variability) occurred in the history under certain measurable conditions, it is likely to recur under similar circumstances. We use the historical \acs{GHI} measurements collected at EPFL by using a pyranometer, which has a temporal resolution of $\qty{50}{ms}$ for the coordinate in \Cref{tab-simulation_setup}. The training phase is performed offline to reduce \acs{RT} computational complexity, where k-means clustering algorithm is used to partition the historical observations into $k$ distinct clusters. During \acs{RT} operation, the model identifies the representative cluster by finding the centroid with the minimum distance to the latest pattern and reports the 10th and 90th percentiles.

\subsubsection{Medium-term GHI forecasting}
For real-world deployment, \acs{GHI} forecasts are obtained from the external numerical weather prediction service provided by MeteoSwiss \cite{NumericalWeatherForecasting}. Specifically, the ICON-CH1-EPS model is used, which predicts the future evolution of atmospheric conditions over the Alpine region with a grid resolution of approximately $\qty{1}{km}$, 1 control member and 10 perturbed ensemble members, forecast horizon of 33~hours, temporal resolution of 1~hour and updates every 3~hours. The grid point closest to the geographical location mentioned above is selected, and the hourly forecasts are interpolated to a temporal resolution of 15~minutes.

\subsection{Carbon intensity forecasting}

Predictions of the carbon intensity of the public power grid are required to account for the emission costs associated with imported electricity and to enable carbon-aware operation of the system. Carbon intensity estimation and accounting can be performed at the transmission-system level or at a more granular regional level \cite{liuCarbonFlowTracing2027}. In this work, the carbon intensity forecasts of the Swiss transmission grid are obtained from the Electricity Maps service \footnote{In alignment with academic sandbox protocols of Electricity Maps for structural validation, a standard randomization factor is present in the carbon signal values; however, the underlying temporal and structural trends remain representative for evaluating the control framework's optimization trajectories.} \cite{electricitymaps2025data}.

\subsection{Intraday Price Forecasting} \label{sec-app-forecasting-id-price}

Although the simulated system considered in this simulation study operates at a relatively small scale, it is experimentally designed to explore the possibility of participating in the \acs{ID} electricity market. This participation allows the system to adjust its \acs{DA} dispatch schedule and thereby reduce potential imbalance penalties. Consequently, forecasts of \acs{ID} electricity prices are required. 

The continuous \acs{ID} electricity market remains relatively underexplored in the literature, particularly with respect to price forecasting \cite{scholzPredictionElectricityPrices2021}. Existing studies have primarily focused on trading strategies or final price estimation. For instance, \cite{aidOptimalTradingProblem2016} proposed an optimal \acs{ID} trading strategy based on a continuous-time stochastic price model, while \cite{UNIEJEWSKI20191533} introduced a LASSO-based approach for forecasting the final \acs{ID} price which identified the most recent \acs{ID} price and the corresponding \acs{DA} price that corresponds to the same hour as the most influential explanatory variables. Moreover, the recent study \cite{yuOrderbookFeatureLearning2026} shows that the \acs{ID} price forecasting might be tractable if the orderbook is available.

In this work, the \acs{ID} spot market of the Swiss bidding zone is selected. Historical data for continuous \acs{ID} trading prices, including low, average, and high prices, are obtained from \cite{energy-chartsEnergyChartsAPI}, which records historical final prices for hourly products. Only one-hour delivery products are considered, as the trading volumes of semi-hourly and quarter-hourly products are limited, leading to frequent zero-price records in the dataset \cite{energy-chartsEnergyChartsAPI}. A linear regression framework is adopted to forecast the final \acs{ID} prices, including the low, average, and high values, as well as their associated confidence intervals. The following explanatory variables are used:
\begin{itemize}
\item final \acs{ID} prices lagged by 1~hour, 1~day, and 1~week,
\item Swiss \acs{DA} electricity price and residual load forecast reported by \cite{entsoeTransparencyPlatform}.
\end{itemize}

\printcredits

\section*{Acknowledgment}

This work is supported by École polytechnique fédérale de Lausanne, through the Heating Bits Solutions for Sustainability (S4S) project.

\bibliographystyle{ieeetr}

\bibliography{RT-DTC-MPC}



\end{document}